\documentclass[preprint, 11pt]{elsarticle}
\usepackage[utf8]{inputenc}
\usepackage{graphicx, wrapfig, subcaption, setspace, float}
\usepackage{amsfonts}
\usepackage[T1]{fontenc}
\usepackage{etoolbox}
\usepackage{amsthm}
\usepackage[table]{xcolor}
\usepackage{bbold}
\usepackage{enumerate}
\usepackage{geometry}
\usepackage{mathtools}
\usepackage{enumitem}
\usepackage{caption}
\usepackage{parskip}
\usepackage{booktabs}
\usepackage{threeparttable}
\usepackage{makecell}
\graphicspath{{./figs/}}
\usepackage{textgreek}
\usepackage{amssymb}
\usepackage{relsize}
\usepackage{todonotes}

\usepackage[unicode]{hyperref}
\definecolor{ForestGreen}{rgb}{0.15,0.416,0.18}
\definecolor{EgyptBlue}{rgb}{0.063,0.2,0.65}
\hypersetup{
  colorlinks=true
}

\makeatletter
\def\ps@pprintTitle{%
  \let\@oddhead\@empty
  \let\@evenhead\@empty
  \def\@oddfoot{\reset@font\hfil\thepage\hfil}
  \let\@evenfoot\@oddfoot
}
\makeatother

\newcommand{\reals}[0]{\ensuremath{\mathbb{R}}}

\newcommand{\domain}[0]{\ensuremath{\mathcal{D}}}
\newcommand{\rzero}[0]{\ensuremath{\mathcal{R}_0}}
\newcommand{\xvec}[0]{\ensuremath{\mathbf{x}}}

\newcommand{\Seq}[0]{\ensuremath{S^{*}}}
\newcommand{\Ieq}[0]{\ensuremath{I^{*}}}
\newcommand{\Req}[0]{\ensuremath{R^{*}}}
\newcommand{\Weq}[0]{\ensuremath{W^{*}}}
\newcommand{\Jeq}[0]{\ensuremath{J^{*}}}

\newcommand{\Wupper}[0]{\ensuremath{\overline{W}}}
\newcommand{\Wlower}[0]{\ensuremath{\underline{W}}}

\newcommand{\BetaCrit}[0]{\ensuremath{\tilde{\beta}}}
\newcommand{\SuperCond}[0]{\ensuremath{\Theta}}

\newcommand{\jac}[0]{\ensuremath{\mathbf{J}}}
\newcommand{\sgn}[0]{\ensuremath{\mathrm{sign}}}
\newcommand{\gh}[0]{\ensuremath{\mathrm{GH}}}
\newcommand{\h}[0]{\ensuremath{\mathrm{H}}}

\newcommand{\SuperBound}[0]{\ensuremath{\mathbf{b}}}

\newcommand{\matcont}[0]{\texttt{MatCont}}

\newtheorem{lemma}{Lemma}
\newtheorem{theorem}{Theorem}

\begin{document}

\begin{frontmatter}
\title{Bifurcation analysis of waning-boosting epidemiological models with repeat infections and varying immunity periods}

\author[szte]{R.~Opoku-Sarkodie} 

\author[szte,NatLab]{F.A.~Bartha\corref{cor1}}
\ead{barfer@math.u-szeged.hu}

\author[szte,NatLab]{M.~Polner}

\author[szte,NatLab]{G.~R\"ost}

\cortext[cor1]{Corresponding author}

\address[szte]{{Bolyai Institute, University of Szeged},
            {Aradi v\'ertan\'uk tere 1, Szeged}, 
            {H-6720}, 
            {Hungary}}
            
\address[NatLab]{{National Laboratory for Health Security}, 
        {Aradi v\'ertan\'uk tere 1, Szeged}, 
        {H-6720}, 
        {Hungary}}

\begin{abstract}
We consider the SIRWJS epidemiological model that includes the waning and boosting of immunity via secondary infections. We carry out combined analytical and numerical investigations of the dynamics. The 
formulae describing the existence and stability of equilibria are derived. Combining this analysis with numerical continuation techniques, we construct global bifurcation diagrams with respect to several epidemiological parameters. The bifurcation analysis reveals a very rich structure of possible global dynamics. We show that backward bifurcation is possible at the critical value of the basic reproduction number, $\rzero = 1$. Furthermore, we find stability switches and Hopf bifurcations from steady states forming multiple endemic bubbles, and saddle-node bifurcations of periodic orbits. Regions of bistability are also found, where either two stable steady states, or a stable steady state and a stable periodic orbit coexist. This work provides an insight to the rich and complicated infectious disease dynamics that can emerge from the waning and boosting of immunity. 
\end{abstract}

\begin{keyword}
  waning immunity \sep immune boosting \sep SIRWJS system \sep bifurcation diagram \sep backward bifurcation \sep Hopf bifurcation
 \end{keyword}
\end{frontmatter}

\section{Introduction}
\label{sec:intro}
Compartmental models based on the Susceptible-Infectious-Recovered ($SIR$) framework, have been used to study the transmission dynamics of infectious diseases in a population. The classical $SIR$ model assumes lifelong and perfect immunity upon recovery from the infection.
An extension of the $SIR$ model, known as the Susceptible-Infectious-Recovered-Susceptible ($SIRS$) model, accounts for the loss of immunity and can capture the long term persistence of diseases in a population. However, it is unable to reproduce oscillatory dynamics, which has been frequently experienced in real life.

Through the addition of a $W$ compartment, the Susceptible-Infectious-Recovered-Waned-Susceptible ($SIRWS$) model can incorporate both the waning and boosting of immunity. Individuals from the $R$ compartment, after the some time, move to the $W$ compartment where they have less immunity than the recovered class $R$, but still more immunity than the fully susceptible class $S$. Moreover, when an individual is in the $W$ compartment, and exposed to the pathogen again, then their immunity can be boosted which can be modeled by moving back to the highly immune $R$ compartment from $W$, without experiencing the infected state. The $SIRWS$ model already exhibits a surprisingly rich dynamics with three distinct features depending on the degree of boosting — fixed points, limit cycles, and bistability between the two. For a comprehensive study of waning and boosting of immunity in a very general setting, we refer to Barbarossa et al. \cite{general}. 

Several authors have extended the $SIRWS$ model to explore additional questions, such as the role of age structure, vaccination, seasonal forcing, and strain dynamics. Carlsson et al. \cite{carlsson} and Lavine et al. \cite{lavine2011natural} examined the resurgence of pertussis by extending the $SIRWS$ model to include age-structure and vaccination. The impact of waning and boosting of immunity on COVID-19 dynamics was studied using an age structured model in \cite{childs}. Leung et al. \cite{leung2018infection} showed that the relative duration of vaccine-induced immunity and infection-induced immunity plays a significant role in determining epidemiological dynamics. Dafilis et al. \cite{dafilis2012influence} considered seasonal forcing of disease transmission and found highly unpredictable behavior. Further work considered the interaction of similar pathogens and demonstrated the interesting behavior when two phenomena that can cause oscillations — strain dynamics with cross-immunity and waning/boosting of immunity — are coupled. 

A common feature of the previous $SIRWS$-models is the assumption of identical expected transition times from $R$ to $W$ and thereon from $W$ to $S$. In our previous work \cite{Richmond1}, we have investigated the effects of breaking this symmetry, {\it i.e.} we considered arbitrary partitioning of the total immune period (the overall expected transition time from $R$ to $S$) between the $R$ and the $W$ states. We found that the modified model exhibits rich dynamics and displays additional complexity with respect to the symmetric partitioning. 

This article presents an extension of the $SIRWS$ model where boosting of immunity occurs strictly via undergoing a secondary infection period, by inserting an additional compartment $J$ from $W$ to $R$. Such an extended system was already studied by Strube et al. \cite{strube} permitting, in addition, allowing immune boosting directly from $W$ to $R$ for a fraction of the cases. We do not consider this latter possibility here, only the boosting via $J$. However, \cite{strube}, similarly to \cite{dafilis2012influence, leung2018infection}, assumed identical transition times from $R$ to $W$ and $W$ to $S$. In contrast, here we investigate how an asymmetric partitioning of the total immune period affects the dynamics, enabling additional bifurcations.

We determine the stability of the endemic equilibria and analyze the parameter regimes in which fixed points, limit cycles, and bistability occur. We establish the possibility of a backward transcritical bifurcation at $\rzero = 1$. Our analysis leads to very complicated dynamics and convoluted bifurcation diagrams. 
The results have implications for the control and prevention of infectious diseases and highlight the need for continued research in this area, to understand long term disease dynamics in populations.

\section{Description of the SIRWJS model: 
a compartmental model with waning and boosting, where secondary exposure can make the host infective}
\label{sec:sirwjs}
In this section, we describe the SIRWJS model, 
which incorporates a secondary infectious state, labelled $J$, via which boosting of immunity occurs. 
Primarily, the SIRWJS model consists of the following compartments: 
those who are susceptible ($S$) to the infection may become infected ($I$) upon adequate contact with an infectious individual. The recovered population is further divided into two 
compartments based on their level of immunity. Upon recovery from $I$, individuals move to $R$ having full immunity. Later, their immunity may weaken and they progress to the $W$ compartment representing waning immunity. Upon re-exposure to the pathogen, members of $W$ move into the 
$J$ compartment representing secondary infections.
These individuals eventually recover from the secondary infection and transition back to $R$ where hosts are fully immune. The path from $W$ to $R$ results in 
a boosting of the individual's immunity level. 
On the other hand, in the absence of re-exposure to the disease causing pathogen, hosts eventually lose their immunity modelled as a transition from $W$ back
to the $S$ compartment where 
they are fully susceptible again to the infection.

\begin{figure}[H]
	\centering
	\includegraphics[scale=0.35]{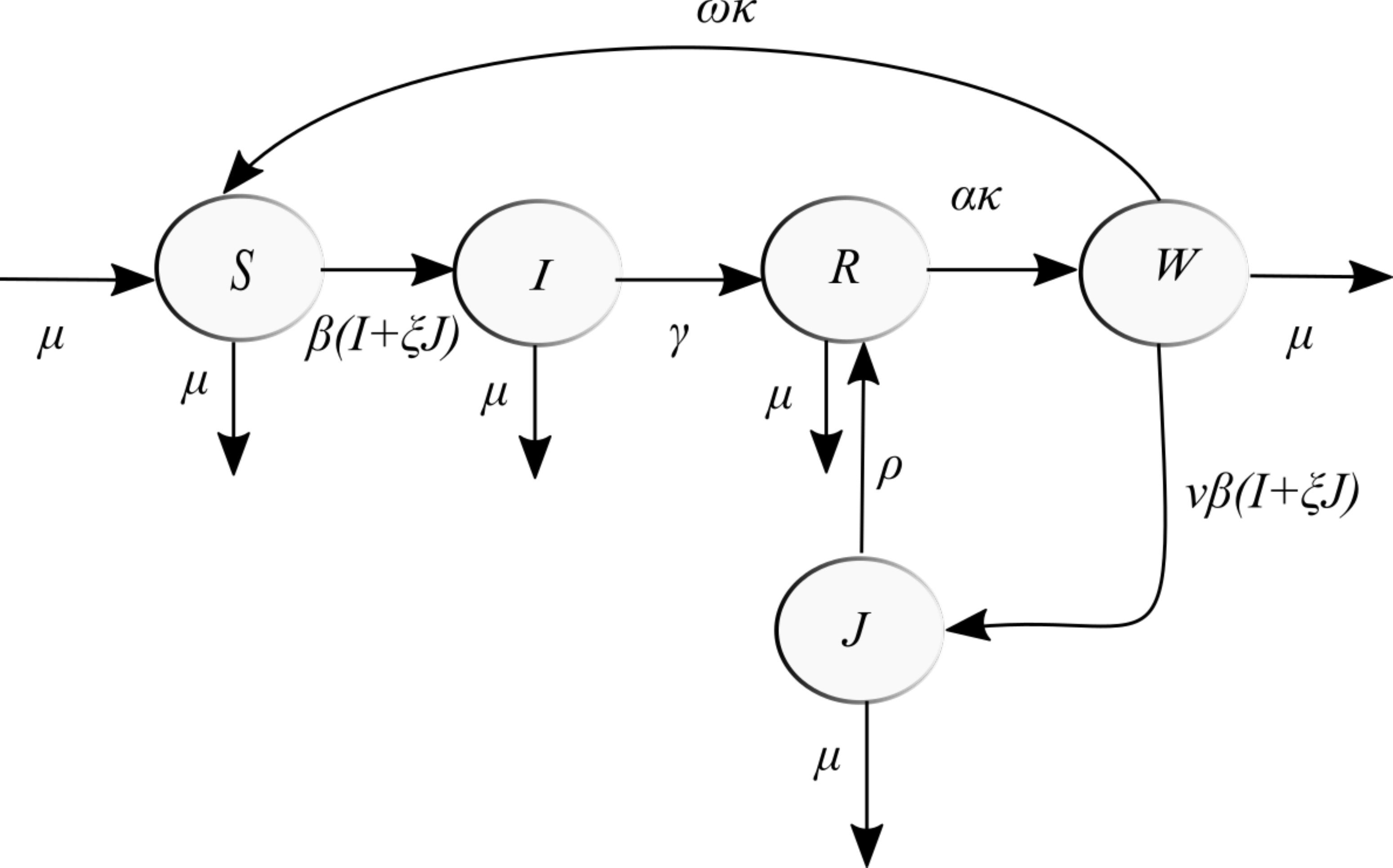}
	\caption{Flow diagram for the SIRWJS system with sub-clinical state.}\label{fig:chart1}
\end{figure}

Figure \ref{fig:chart1} shows the flow chart of the SIRWJS system, where boosting occurs via $J$. The primary 
force of infection is $\beta (I + \xi J)$, 
where $\xi$ is the infectivity of secondary infection 
relative to primary infection and 
$\beta$ is the transmission rate. Thus, both $I$ and $J$ are infectious compartments, and individuals in these compartments can infect susceptibles and also boost a waning immunity. 
The death rate, $\mu$, is assumed to be the same as the birth rate, 
$\gamma$ and $\rho$ are the recovery rates from 
the primary and secondary infections respectively, 
while $\kappa$ is the immune decay rate. 
Boosting of immunity occurs via the $J$ compartment 
using the boosting coefficient $\nu$.

Many previous waning-boosting models assumed that the average time spent in $R$ and $W$ compartments are the same. Here, following \cite{Richmond1}, we relax this restrictive assumption of symmetric partition of the immunity period, by introducing two additional parameters 
$\alpha>1$ and $\omega>1$, such that the time spent in 
$R$ is $1/(\alpha\kappa)$ and the time spent in 
$W$ is $1/(\omega\kappa)$. 
Then, the total period of immune protection is
\begin{equation}\label{eq:immune_protection}
  \frac{1}{\alpha \kappa} + \frac{1}{\omega \kappa} = 
  \frac{1}{\kappa},
\end{equation} 
under the assumption of 
$\alpha + \omega = \alpha \omega$. Note that the formulation of similar models in earlier works such as \cite{dafilis2012influence} is equivalent with the restriction of parameters $\alpha=\omega=2$.

The descriptions and assumptions on the system parameters are summarized 
in Table~\ref{table:parameter_assumptions}.
\begin{table}[H]
\begin{center}
\begin{tabular}{ |c|c|c| } 
\hline
$\beta$ & $ > 0$ & transmission rate \\ 
$\xi$ & $ \geq 0$ & \makecell{relative infectivity of secondary infections\\ with respect to primary} \\ 
\hline
$\mu$ & $ > 0$ & birth and death rate \\ 
\hline
$\gamma$ & $ > 0$ & recovery rate from primary infection \\ 
$\rho$ & $ > 0$ & recovery rate from secondary infection \\ 
\hline
$\kappa$ & $ > 0$ & immune decay rate \\ 
\hline
$\alpha^{-1}$ & $ \in (0, 1)$ & \makecell{relative size of the first 
immune \\ protection period from $R \longrightarrow W$} \\ 
$\omega^{-1}$ & $(1 - \alpha^{-1})  \in (0, 1)$ & 
\makecell{relative size of the second 
immune \\ protection period from $W \longrightarrow S$} \\ 
\hline
$\nu$ & $ > 0$ & boosting coefficient \\ 
\hline
\end{tabular}
\end{center}
\caption{Parameters of the SIRWJS system.}
\label{table:parameter_assumptions}
\end{table}
\noindent
We consider all parameters to be 
positive, but $\xi$ is allowed to take value zero as well. The case $\xi = 0$ 
represents the scenario when people in secondary infection are not infectious, whilst 
$\xi = 1$ describes the scenario when the secondary infection 
is equally infectious to the primary infection. We may allow $\xi > 1$ modeling 
reinfections that are more severe than the primary. 

We now formulate the governing system of 
ordinary differential equations describing the 
dynamics presented in Figure~\ref{fig:chart1} as
\begin{equation}
\label{eq:sirw} 
	\begin{split} 
	\frac{dS}{dt} & = \, -\beta (I + \xi J) S + 
	    \omega \kappa W + \mu(1 - S), \\
	\frac{dI}{dt} & = \,  \beta (I + \xi J) S - 
	    \gamma I - \mu I, \\
	\frac{dR}{dt} & = \, \gamma I - \alpha \kappa R + 
	    \rho J - \mu R, \\
	\frac{dW}{dt} & = \, \alpha \kappa R - 
	    \omega \kappa W - \nu \beta (I + \xi J) W - 
	    \mu W, \\
	\frac{dJ}{dt} & = \,  \nu \beta (I + \xi J) W -
	    \mu J - \rho J. 
	\end{split}
\end{equation}
System \eqref{eq:sirw} models a constant size population that we normalized to $1$. Thus, our primary interest is 
in non-negative solutions satisfying $S + I + R + W + J = 1$ 
for all $t$. These solutions we refer to as 
\emph{epidemiologically feasible}. 

Using the substitution 
\begin{equation}
\label{eq:Rsubstitution}
    R = 1 - S - I - W - J,
\end{equation}
we get the reduced system 
\begin{subequations}\label{eq:sirw_reduced}
	\begin{align} 
	\frac{dS}{dt} & =\, - \beta (I + \xi J) S + 
	    \omega \kappa W + \mu(1 - S), 
	    \label{eq:sirw_reduced_S} \\
	\frac{dI}{dt} & = \,  \beta (I + \xi J) S -
	    \gamma I - \mu I, 
	    \label{eq:sirw_reduced_I} \\
	\frac{dW}{dt} & = \, \alpha \kappa (1 - S - I - W - J) - 
	    \omega \kappa W - \nu \beta (I + \xi J) W -
	    \mu W,
	    \label{eq:sirw_reduced_W} \\
	\frac{dJ}{dt} & = \, \nu \beta (I + \xi J) W -
	    \mu J - \rho J. 
	    \label{eq:sirw_reduced_J}
	\end{align}
\end{subequations}
\noindent
Note that the feasible region for our epidemiological setting 
\begin{equation*}
    (S(t), I(t), W(t), J(t)) \in \domain := 
    \left\{ (s, i, w, j) \in \reals_{\geq 0}^4 ~ | ~ 0 \leq s + i + w + j \leq 1 \right\} 
\end{equation*}
is forward invariant. This follows from the observation that the components in \eqref{eq:sirw_reduced} remain non-negative, and then
\begin{equation*}
    \frac{d(S + I + W + J)}{dt} = 
    (\alpha \kappa + \mu) (1 - S - I - W - J) - 
    \gamma I - \rho J
\end{equation*}
indicates that the sum cannot exceed $1$.

\section{Equilibria and stability analysis}
\label{sec:equilibria}
Now we turn our attention to finding equilibria 
$(\Seq, \Ieq, \Req, \Weq, \Jeq)$ of \eqref{eq:sirw}. 
The following lemma establishes that  
feasible ones 
arise from non-negative steady states 
of the reduced system \eqref{eq:sirw_reduced}.

\begin{lemma}\label{lemma:relevance}
Let $(\Seq, \Ieq, \Weq, \Jeq)$ be a non-negative equilibrium 
of \eqref{eq:sirw_reduced}. Then
\begin{equation*}
    (\Seq, \Ieq, \Weq, \Jeq) \in \domain
\end{equation*}
and, hence, $(\Seq, \Ieq, \Req, \Weq, \Jeq)$ with 
$\Req := 1 - \Seq - \Ieq - \Weq - \Jeq$ is 
epidemiologically feasible.
\end{lemma}

\begin{proof}
Equilibria of \eqref{eq:sirw_reduced} are obtained as 
solutions of 
\begin{subequations}\label{eq:sirw_reduced_equlibria}
	\begin{align} 
	- \beta (\Ieq + \xi \Jeq) \Seq + 
	    \omega \kappa \Weq + \mu (1 - \Seq) & = \, 0, 
	    \label{eq:sirw_reduced_equlibria_S} \\
	  \beta (\Ieq + \xi \Jeq) \Seq - 
	    \gamma \Ieq - \mu \Ieq & = \, 0, 
	    \label{eq:sirw_reduced_equlibria_I} \\
	\alpha \kappa (1 - \Seq - \Ieq - \Weq - \Jeq) - 
	    \omega \kappa \Weq - 
	    \nu \beta (\Ieq + \xi \Jeq) \Weq - 
	    \mu \Weq & = \, 0, 
	    \label{eq:sirw_reduced_equlibria_W} \\
	\nu \beta (\Ieq + \xi \Jeq) \Weq - 
	    \mu \Jeq - \rho \Jeq & = \, 0. 
	    \label{eq:sirw_reduced_equlibria_J}
	\end{align}
\end{subequations}
Summing all equations yields
\begin{equation*}
    (\mu + \alpha \kappa) (1 - \Seq - \Ieq - \Weq - \Jeq) - 
    \gamma \Ieq - \rho \Jeq = 0,
\end{equation*}
thus, $\Seq + \Ieq + \Weq + \Jeq \leq 1$ as 
$0 < \gamma, \rho, (\alpha \kappa + \mu)$ and 
$0 \leq \Ieq, \Jeq$.
\end{proof}
\noindent
The converse is readily satisfied, namely, 
given $(\Seq, \Ieq, \Req, \Weq, \Jeq)$ 
an epidemiologically feasible equilibrium of
\eqref{eq:sirw}, we have 
$(\Seq, \Ieq, \Weq, \Jeq) \in \domain$.
Consequently, in the following 
we concentrate on finding non-negative equilibria of 
\eqref{eq:sirw_reduced} and, then, we study 
their local stability. 

Note that 
equation~\eqref{eq:sirw_reduced_equlibria_S}  
implies $\Seq > 0$ for any non-negative equilibrium.

\subsection{Disease free equilibrium}
\label{sec:DFE}
Assume $\Ieq = 0$. Then, $\Jeq = 0$ follows from 
\eqref{eq:sirw_reduced_equlibria_I} and the above observation 
on $\Seq$ being positive. The non-negativity implies 
$\Seq = 1$ and, in turn, $\Weq = 0$ from 
\eqref{eq:sirw_reduced_equlibria_S}. 
The resulting equilibrium 
$(\Seq, \Ieq, \Weq, \Jeq) = (1, 0, 0, 0)$ is referred 
to as the \emph{disease free equilibrium} (DFE).

We note that 
even if we relax the non-negativity condition, no other 
equilibria exists with $\Ieq = 0$. We refer to our computer algebra codes for further details \cite{github}.

\subsection{Existence of non-trivial equilibria}
\label{sec:EE}
Let us now consider $\Ieq \neq 0$ and assume $\xi > 0$. We  
will return to the case $\xi = 0$ later. 
Equation \eqref{eq:sirw_reduced_equlibria_J} implies 
$\Weq = 0$ if and only if $\Jeq = 0$ and then
\begin{equation*}
\begin{split}
    \Seq &= \frac{\gamma + \mu}{\beta}, \\
    \Ieq &= \mu \left( 
        \frac{1}{\gamma + \mu} - \frac{1}{\beta}
    \right), \\
    \alpha \kappa \gamma & \left( 
        \frac{1}{\gamma + \mu} - \frac{1}{\beta} 
    \right) = 0.
\end{split}
\end{equation*}
Thus, in this case $\beta = \gamma + \mu$ and 
we, again, obtain the DFE. Hence, 
we may assume both $\Weq \neq 0$ and $\Jeq \neq 0$. 
Also, $\rho + \mu - \nu \beta \xi \Weq \neq 0$ as 
equality would imply $\Ieq = 0$ 
by \eqref{eq:sirw_reduced_equlibria_J}. 

After these preliminary observations, 
we begin by expressing $(\Seq, \Ieq, \Jeq)$ 
in terms of $\Weq$. 
From \eqref{eq:sirw_reduced_equlibria_I} and 
\eqref{eq:sirw_reduced_equlibria_J}, we obtain 
\begin{equation*}
    \frac{\Jeq}{\Ieq} = 
    \frac{\gamma + \mu - \beta \Seq}{\beta \xi \Seq} = 
    \frac{\nu \beta \Weq}{\rho + \mu - \nu \beta \xi \Weq}
\end{equation*}
yielding
\begin{equation}\label{eq:Sstar}
    \Seq = \frac{\gamma + \mu}{\beta} - 
        \frac{\nu \xi (\gamma + \mu)}{\rho + \mu} \Weq.
\end{equation}
Then, adding \eqref{eq:sirw_reduced_equlibria_S} and 
\eqref{eq:sirw_reduced_equlibria_I} results in 
\begin{equation*}
    \Ieq = \frac{\omega \kappa \Weq + 
        \mu (1 - \Seq)}{\gamma + \mu}
\end{equation*}
that simplifies to 
\begin{equation}\label{eq:Istar}
    \Ieq = \mu \left( 
        \frac{1}{\gamma + \mu} - \frac{1}{\beta}
    \right)  + \left(
        \frac{\omega \kappa}{\gamma + \mu} + 
        \frac{\mu \nu \xi}{\rho + \mu}
    \right) \Weq.
\end{equation}
Finally, \eqref{eq:sirw_reduced_equlibria_J} and 
\eqref{eq:Sstar} gives 
\begin{equation}\label{eq:Jstar}
    \Jeq = \frac{\nu \beta \Ieq \Weq}{\rho + \mu 
        - \nu \beta \xi \Weq}.
\end{equation}
We note that \eqref{eq:Jstar} could be expanded 
solely in terms of $\Weq$ using \eqref{eq:Istar}. Nevertheless, 
the added complexity would serve no benefit and, thus, 
the expansion is omitted. 

Using the above formulae,  we obtain a 
quadratic equation for $\Weq$ from 
\eqref{eq:sirw_reduced_equlibria} as 
\begin{equation}\label{eq:Wstar}
    f(\Weq) := A (\Weq)^2 + B \Weq + C = 0,
\end{equation}
with
\begin{align}\label{eq:quad_eq_coeff}
\begin{split}
A &= \nu \beta^2 \Big[
    - \nu \xi^2 (\gamma +\mu ) Q_0 +
    \xi Q_1 + 
    \big( \alpha \kappa (\rho + \mu) - 
     \nu \xi \mu (\gamma + \mu) - 
     \eta \kappa (\rho + \mu) \big) Q2
    \Big], \\
B &= \beta (\rho + \mu) \Big[ 
    \big(
     \nu \xi (\gamma + \mu) - 
     \nu \xi (\beta - \gamma - \mu)
    \big) Q_0 - 
    Q_1 - 
    \nu \mu (\beta - \gamma - \mu) Q_2
    \Big] ,\\
C &= (\beta -\gamma -\mu ) (\rho + \mu)^2 Q_0, 
\end{split}
\end{align}
where
\begin{align}\label{eq:Qnotations}
\begin{split}
    \eta &:= \alpha + \omega = \alpha \omega, \\
    Q_0 &:= \alpha \kappa \gamma, \\
    Q_1 &:= \left[
        (\gamma + \mu) (\eta \kappa + \mu) + 
        \eta \kappa^2
    \right] (\rho + \mu), \\
    Q_2 &:= \alpha \kappa + \rho + \mu.
\end{split}
\end{align}
Therefore, based on the sign of the discriminant
$\Delta = B^2 - 4 A C$, system 
\eqref{eq:sirw_reduced_equlibria} has $0$, $1$ or $2$ 
additional real solutions besides the DFE. 
Note that an equilibria originating 
from a real root of the quadratic equation 
coincides with the DFE if and only if the root is zero.

Let us now investigate the non-negativity 
of these non-trivial equilibria. Based on our 
initial considerations at the beginning of this 
section, we are looking for positive 
solutions and, thus, we assume 
that \eqref{eq:Wstar} has a solution $\Weq > 0$. 
Then, the inequality 
\begin{equation}\label{eq:wUpper}
    \Weq < \frac{\rho + \mu}{\nu \beta \xi} =: \Wupper
\end{equation}
must hold in order to ensure $\Seq >0$ based 
on \eqref{eq:Sstar}. Similarly, 
\begin{equation}\label{eq:wLower}
    \Weq > \frac{\mu(- \beta + \gamma + \mu)(\rho + \mu)}{
        \beta(\mu \nu \xi (\gamma + \mu) + 
        \omega \kappa (\rho + \mu) )} =: \Wlower
\end{equation}
follows from \eqref{eq:Istar}. 
Finally, one can see from \eqref{eq:Jstar} that $\Jeq > 0$ 
readily follows from \eqref{eq:wUpper}, \eqref{eq:wLower}, 
and $\Weq > 0$. Summarizing these findings and using 
Lemma~\ref{lemma:relevance} yield that 
a solution $\Weq$ of \eqref{eq:Wstar} leads to 
an epidemiologically feasible equilibrium other than the DFE  
by \eqref{eq:Rsubstitution}, \eqref{eq:Sstar}, \eqref{eq:Istar}, and \eqref{eq:Jstar} if and only if 
\begin{equation}
\label{eq:Wbounds}
    \max\{0, \Wlower\} < \Weq < \Wupper.
\end{equation}
Note that the above conditions guarantee 
the non-negativity 
of the equilibrium, hence, it follows from  
Lemma~\ref{lemma:relevance} that 
$(\Seq, \Ieq, \Weq, \Jeq) \in \domain$. 
In particular, $\Weq \leq 1$ must hold implying that 
no such $\Weq$ may exist if $\Wlower \geq 1$. 

For the upper bound, straightforward calculation shows that the quadratic formula \eqref{eq:Wstar} is 
negative at
\begin{equation}
\label{eq:fAtUpper}
    f(\overline W) = - 
    \frac{(\rho + \mu)^2 (\mu \nu \beta \xi + 
    (\eta - \alpha) \kappa (\rho + \mu))}{\nu \xi^2}Q_2 < 0,
\end{equation} 
given any parametrization conforming Table~\ref{table:parameter_assumptions}. 

Let us now analyze the lower bound and the sign of 
$f$ at that point. Clearly, $0 \geq \Wlower$ if and only if 
$\beta \geq \gamma + \mu$. Note that the 
\emph{basic reproduction number} $\rzero$ of the system 
\eqref{eq:sirw_reduced} -- and of \eqref{eq:sirw} --
is obtained as the spectral radius of 
\begin{equation*}
\begin{split}
    - \mathbf{T} \mathbf{\Sigma}^{-1} &= 
    - \left[
    \begin{array}{ccc}
    \beta & 0 & \beta \xi \\
    0 & 0 & 0 \\
    0 & 0 & 0 \\
  \end{array}\right]
  \times
  \left[
    \begin{array}{ccc}
    - (\gamma + \mu) & 0 & 0 \\
    -\alpha \kappa & -(\alpha \kappa + \omega \kappa + \mu) & -\alpha \kappa \\
    0 & 0 & - (\rho + \mu) \\
  \end{array}\right]^{-1}  \\
  &= \left[
    \begin{array}{ccc}
    \frac{\beta}{\gamma + \mu} & 0 & \frac{\beta \xi}{\gamma + \mu} \\
    0 & 0 & 0 \\
    0 & 0 & 0 \\
  \end{array}\right]
\end{split}
\end{equation*}
via the next generation matrix method \cite{diekmann}, where $\mathbf{T}$ and 
$\mathbf{\Sigma}$ represent the transmission part describing
the production of new infections, and the transition 
part describing changes in state, of the linearized infected subsytem composed of $(I, W, J)$, 
respectively, where $\mathbf{T}+ \mathbf{\Sigma}$ is the corresponding Jacobian. Therefore,
\begin{equation*}
    \rzero = \frac{\beta}{\gamma + \mu} 
\end{equation*}
and the condition $\beta \geq \gamma + \mu$  translates to 
$\rzero \geq 1$.

Consider now $\rzero > 1$. The $y$-intercept of the parabola 
in \eqref{eq:Wstar} is positive, i.e., 
$f(0) = C > 0$. 
Hence, $f$ has exactly one root in
the interval $(0, \Wupper)$ and, as a consequence, 
\eqref{eq:sirw_reduced} has one other 
epidemiologically feasible equilibrium besides 
the DFE. 
This new equilibrium is referred to as the 
\emph{endemic equilibrium} (EE). 
Note that, independent of the parametrization, 
the formula for EE is obtained by using the root
\begin{equation}
\label{eq:Wstar-endemic}
    \Weq \equiv \Weq_{-} = 
        \frac{-B - \sqrt{B^2 - 4 A C}}{2 A}.
\end{equation}

The case $\rzero < 1$ is more involved. The lower bound in \eqref{eq:Wbounds} is now given by 
$\Wlower$ and elementary calculations yield
\begin{equation}
\label{eq:fAtLower}
    f(\Wlower) = 
    \frac{
        (\beta - \gamma - \mu) (\rho + \mu)^2 
        (\omega \kappa (\rho + \mu) Q_0 +
        \mu Q_1) 
        (\mu \nu \beta \xi + 
        \omega \kappa (\rho + \mu))}{
        \left(\mu \nu \xi (\gamma + \mu) 
        + \omega \kappa (\rho + \mu) \right)^2}
    < 0.
\end{equation}
Thus, by \eqref{eq:fAtUpper} and \eqref{eq:fAtLower}, 
if $f$ has a root in $(\Wlower, \Wupper)$, then 
$f$ is a downward parabola with 
non-negative discriminant $\Delta$. Moreover, if 
$\Delta > 0$, then it has two roots of 
the sought quality leading to two other 
epidemiologically feasible equilibria. 
A more thorough sign analysis of $\Delta$ 
reveals that if such equilibria exist then they 
do so for an interval of $\beta$ values in the 
left neighbourhood of $\gamma + \mu$ distant from $0$. 

\begin{theorem}
\label{thm:otherEquilibria}
Let
\begin{equation}
\label{eq:SuperCond}
    \SuperCond = \nu \xi (\gamma + \mu) Q_0 - Q_1, 
\end{equation}
with $Q_0, Q_1$ defined in \eqref{eq:Qnotations}. If $\SuperCond > 0$, then there is a 
$0 < \BetaCrit < \gamma + \mu$ such that,
besides the DFE, there are two other 
epidemiologically feasible equilibria for 
$\beta \in (\BetaCrit, \gamma + \mu)$ and only 
the DFE for $\beta < \BetaCrit$. 
On the other hand, if $\SuperCond \leq 0$, 
then the only epidemiologically feasible equilibrium 
is the DFE for $\rzero < 1$.
\end{theorem}

\paragraph{Remark}
We emphasize that the possibility of  
$\SuperCond > 0$ is not a consequence of 
the asymmetric partitioning we consider in this manuscript 
as in the symmetric case it translates to
\begin{equation*}
    \nu \xi (\gamma + \mu) 2 \kappa \gamma - 
    \left[
        (\gamma + \mu) (4 \kappa + \mu) + 
        4 \kappa^2
    \right] (\rho + \mu) > 0
\end{equation*}
that is clearly satisfiable with an appropriate choice of e.g. $\nu$ or $\xi$. Therefore, the associated results are 
applicable to \cite{strube} as well.

\begin{proof}
We consider $\beta \in (0, \gamma + \mu]$ that is 
$\rzero \leq 1$. By the formulae \eqref{eq:Wstar}, 
\eqref{eq:quad_eq_coeff}, \eqref{eq:wUpper}, 
and \eqref{eq:wLower} 
we have that 
$A$, $B$, $C$, $\Delta$, $f$, $\Wlower$, and $\Wupper$ are 
continuous in $\beta$. 
Recall that $f$ is guaranteed to take negative values 
at the endpoints of the interval $[\Wlower, \Wupper]$ 
as seen in \eqref{eq:fAtUpper} and \eqref{eq:fAtLower}. 
Hence, if for a $\beta_0 \in (0, \gamma + \mu]$ 
the parabola $f$ has two roots in $(\Wlower, \Wupper)$
(and consequently the discriminant $\Delta(\beta_0) > 0$), 
then, due to the continuity of all relevant expressions, 
there exists a corresponding maximal sub-interval 
$ (\underline{\beta}, \overline{\beta})$ with 
\begin{equation*}
    (0, \gamma + \mu) \supseteq 
    (\underline{\beta}, \overline{\beta}) 
    \ni \beta_0
\end{equation*} 
such that the two roots persist (and $\Delta(\beta) > 0$) 
for $\beta \in (\underline{\beta}, \overline{\beta})$. 
Clearly, if $0 < \underline{\beta}$, then $\Delta(\underline{\beta}) = 0$ and, analogously, 
$\overline{\beta} < \gamma + \mu$ implies 
$\Delta(\overline{\beta}) = 0$.

From \eqref{eq:quad_eq_coeff}, we see that the 
discriminant $\Delta$, as a function of $\beta$, takes 
the form
\begin{equation*}
    \Delta(\beta) = (\rho + \mu)^2 \cdot \beta^2 \cdot 
    q(\beta),
\end{equation*}
where $q$ is an upward parabola with lead coefficient 
$\nu^2 (\xi Q_0 + \mu Q_2)^2 > 0$. Hence, 
$\Delta(\beta)$ can have at most two zeros in 
$\beta \in (0, \gamma + \mu]$. 

These observations imply that the subset of 
$(0, \gamma + \mu)$ where $f$ has two roots in 
$(\Wlower, \Wupper)$ must have one of the forms:
\begin{center}
\begin{tabular}{lp{8cm}l} 
-
& $\Delta$ has no zeros in $(0, \gamma + \mu)$: 
& $\emptyset$ or $(0, \gamma + \mu)$, \\[0.5em] 
-
& $\Delta$ has one single zero $\BetaCrit$ in $(0, \gamma + \mu)$: 
& $(0, \BetaCrit)$ or $(\BetaCrit, \gamma + \mu)$, \\[0.5em] 
-
& $\Delta$ has two single zeros $\BetaCrit_1, \BetaCrit_2$ in $(0, \gamma + \mu)$: \newline \phantom{$\Delta$} (or a double zero at $\BetaCrit_1 = \BetaCrit_2$) 
& $(0, \BetaCrit_1) \cup (\BetaCrit_2, \gamma + \mu)$.
\end{tabular}
\end{center}

We can rule out the options having $0$ as a left endpoint 
by noting that 
\begin{equation*}
    \lim_{\beta \to 0} \Wlower = \lim_{\beta \to 0} \Wupper = \infty,
\end{equation*}
thus, in a neighbourhood of $0$, 
the inequality $\Wlower > 1$ holds guaranteeing that no
suitable root exists. Therefore, we are left with two 
possible forms $\emptyset$ and $(\BetaCrit, \gamma + \mu)$ with 
$\BetaCrit > 0$ in the latter. 

In order to finish our proof, we now show that 
the sign of $\SuperCond$ determines if $f$ has a 
root in the left neighbourhood of $\beta = \gamma + \mu$.
First, note that the sign of the $y$-intercept of $f$ is
given as $C(\beta) = 0$ when $\beta=\gamma+\mu$ and $C(\beta) < 0$ for 
$\beta < \gamma + \mu$ and that $\Wlower(\beta)= 0$ when $\beta=\gamma + \mu$. 
Next, the discriminant at 
the critical point is 
\begin{equation*}
    \Delta(\beta)\Big\vert_{\beta = \gamma + \mu} = \SuperCond^2  
    (\gamma + \mu)^2 (\rho + \mu)^2.
\end{equation*}
Finally, the slope of the parabola 
$f$ in \eqref{eq:Wstar} at $\Weq = 0$ as a function of $\beta$ is given by
\begin{equation*}
    B(\beta) = \beta (\rho + \mu) \Big[ 
      \SuperCond - (\beta - \gamma - \mu) 
      \nu (\xi Q_0 + \mu Q_2)
    \Big].
\end{equation*}
Clearly, for $\beta = \gamma + \mu$, 
the inequality $\SuperCond > 0$ 
implies that the above slope is positive 
securing the existence of another root of the parabola $f$ in 
$(0, \Wupper)$ as $f(\Wupper) < 0$ holds. Then, by continuity 
and by $\Delta(\beta)\vert_{\beta=\gamma + \mu} > 0$, we have that this root 
persists in an open neighbourhood of $\beta = \gamma + \mu$. 
On the other hand, when $\SuperCond \leq 0$, the slope is 
non-positive in an open left neighbourhood of 
$\beta = \gamma + \mu$, thus, no other root may exist  there as the $y$-intercept is negative. 
\end{proof}

It is apparent that 
$\rzero = 1$  
marks a significant change in the dynamics. 
We analyze the corresponding bifurcation in 
the following section. Not surprisingly, 
the key expression $\SuperCond$ of
Theorem~\ref{thm:otherEquilibria} will appear 
there as well, broadening our understanding of 
its origin.

\subsection{Transcritical bifurcation at \texorpdfstring{$\rzero=1$}{R0 = 1}}
\label{sec:transcritical}
In this section, we analyze the local stability 
of the DFE and its connection with $\rzero$. First, 
let us consider the Jacobian matrix of our SIRWJS system
\eqref{eq:sirw_reduced} 
\begin{equation*}
\jac = \left[
\begin{array}{cccc}
    - \beta (I + \xi J) - \mu & - \beta S & 
    \omega \kappa & - \beta \xi S \\
    \beta (I + \xi J) & \beta S - (\gamma + \mu) & 
    0 & \beta \xi S \\
    - \alpha \kappa & - \nu \beta W - \alpha \kappa &
    - \nu \beta (I + \xi J) - 
        (\alpha \kappa + \omega \kappa + \mu) &  
        - \nu \beta \xi W - \alpha \kappa \\
    0 & \nu \beta W & 
    \nu \beta (I + \xi J) & \nu \beta \xi W - (\rho + \mu)
\end{array}
\right].
\end{equation*}
and evaluate at the DFE to obtain
\begin{equation*}
    \jac|_{(1, 0, 0, 0)} = 
    \left[
    \begin{array}{cccc}
    - \mu & - \beta & 
    \omega \kappa & - \beta \xi \\
    0 & \beta - (\gamma + \mu) & 
    0 & \beta \xi \\
    - \alpha \kappa & - \alpha \kappa & 
    - (\alpha \kappa + \omega \kappa + \mu) & 
        - \alpha \kappa \\
    0 & 0 & 
    0 & - (\rho + \mu)
  \end{array}\right].
\end{equation*}
The corresponding eigenvalues are
\begin{equation}
\label{eq:eigenvalues_DFE}
    \lambda_{1} =   \beta - (\gamma + \mu), \quad 
    \lambda_{2} = - (\alpha \kappa + \mu), \quad
    \lambda_{3} = - (\omega \kappa + \mu), \quad
    \lambda_{4} = - (\rho + \mu).
\end{equation}
Then, as the eigenvalues 
$\lambda_2$, $\lambda_3$, and $\lambda_4$ are negative 
and $\lambda_1 < 0$ if and only if $\beta < \gamma + \mu$, 
we can conclude that the DFE is locally asymptotically stable 
when $\rzero < 1$ and unstable if $\rzero > 1$. 

The following theorem establishes that a 
transcritical bifurcation happens at $\rzero = 1$. 
We show that the sign of $\SuperCond$, defined in 
\eqref{eq:SuperCond}, gives the direction of this bifurcation. 
The proof relies on Theorem~4.1 of 
\cite{castillo2004dynamical}. 

\begin{theorem}
\label{thm:transcrit}
If $\SuperCond > 0$, then a transcritical bifurcation 
of backward type occurs at $\rzero = 1$, and 
when $\SuperCond < 0$, then a 
transcritical bifurcation of forward type 
occurs at $\rzero = 1$.
\end{theorem}

\begin{proof}
We apply Theorem~4.1 of \cite{castillo2004dynamical} 
to the system $\dot{\xvec} = g(\xvec, b)$, 
where the vector field
\begin{equation*}
    g= (g_{S}, g_{I}, g_{W}, g_{J})
\end{equation*}
is obtained by applying the substitutions for our bifurcation
parameter  
$\beta \to b + \beta^*$ with $\beta^{*} = \gamma + \mu$, 
corresponding to the critical case $\rzero = 1$, and
for the state variables 
$(S, I, W, J) \to (x_S, x_I, x_W, x_J) + (1, 0, 0, 0)$ 
which are then written as 
\begin{equation*}
    \xvec = (x_S, x_I, x_W, x_J).
\end{equation*}

Then, 
$M := D_\xvec g(\mathbf{0}, 0)$ equals to the 
Jacobian matrix of \eqref{eq:sirw_reduced} at the DFE, 
namely to $\jac|_{(1, 0, 0, 0)}$ with $\beta = \beta^*$.
Hence, $M$ has one simple zero eigenvalue and 
three eigenvalues with negative real part 
as in \eqref{eq:eigenvalues_DFE}. 
Now, we calculate the right and left eigenvectors 
$w, v$ of $M$ corresponding to the zero eigenvalue. 
The system $M w = 0$ is underdetermined, 
so we may fix $w_{I} = 1$. Then, 
\begin{equation*}
    w_{S} = - \frac{
        Q_1}{
        (\alpha \kappa + \mu)(\kappa \omega + \mu)
        (\rho + \mu)}, \quad 
    w_{I} = 1, \quad 
    w_{W} = \frac{Q_0}{
        (\alpha \kappa + \mu)(\kappa \omega + \mu)}, \quad 
    w_{J} = 0.
\end{equation*}
Similarly, setting $v_I = 1$ yields 
\begin{equation*}
    v_S = 0, \quad 
    v_I = 1, \quad 
    v_W = 0, \quad 
    v_J = \frac{\xi (\gamma + \mu)}{\rho + \mu}.
\end{equation*}
Now, we need to calculate the following quantities 
\begin{align*}
    Z_1 &= \sum_{k, i, j \in \{S, I, W, J\}} v_{k} w_{i} w_{j} 
        \frac{\partial^{2} g_{k}}{
            \partial x_{i} \partial x_{j}}
            (\mathbf{0}, 0) \qquad \mbox{and} \\ 
    Z_2 &= \sum_{k, i \in \{S, I, W, J\}} v_{k} w_{i} 
        \frac{\partial^{2} g_{k}}{\partial x_{i} \partial b}
            (\mathbf{0}, 0).
\end{align*}
Since $v_S = v_W = 0$, the partial derivatives of 
$g_{S}$ and $g_{W}$ have no influence on the above 
expressions. Also, as $w_{J} = 0$, partial derivatives 
with respect to $x_J$ can be omitted. Thus, we 
are left with the following relevant nonzero second order 
partial derivatives
\begin{equation*}
    \frac{\partial^{2} g_{I}}{
        \partial x_S \partial x_I}(\mathbf{0}, 0) = \beta^*, \quad 
    \frac{\partial^{2} g_{J}}{
        \partial x_I \partial x_W}(\mathbf{0}, 0) = \nu \beta^*, \quad
    \frac{\partial^{2} g_{I}}{
        \partial x_I \partial \beta}(\mathbf{0}, 0) = 1
\end{equation*}
leading to the simplified expressions 
\begin{align*}
    Z_1 &= 2 v_{I} w_{S} w_{I} 
            \frac{\partial^{2} g_{I}}{
            \partial x_S \partial x_I}
            (\mathbf{0}, 0) +
           2 v_{J} w_{I} w_{W}
           \frac{\partial^{2} g_{J}}{
           \partial x_I \partial x_W} 
           (\mathbf{0}, 0) \\[5pt]
    &= \frac{2 \beta^*}{
        (\alpha \kappa + \mu)
        (\kappa \omega + \mu)
        (\rho + \mu)}
        \Big[ 
            \nu \xi (\gamma + \mu) Q_0 - Q_1 
        \Big] \\[5pt]
    &= \frac{2 \beta^*}{
        (\alpha \kappa + \mu)
        (\kappa \omega + \mu)
        (\rho + \mu)} \cdot \SuperCond
        \qquad \qquad \mbox{and} \\[5pt]
    Z_2 &= v_{I} w_{I} 
    \frac{\partial^{2} g_{I}}{
        \partial x_I \partial \beta} 
        (\mathbf{0}, 0) = 
        v_{I} w_{I} = 1.
\end{align*}
As $Z_2 > 0$ for all model parameters, 
only the sign of $Z_1$ decides upon the direction 
of the bifurcation. 
Therefore, if $\SuperCond > 0$ ($< 0$), then 
a transcritical bifurcation of backward (forward) 
type occurs at $\rzero = 1$. 
\end{proof}

\begin{figure}[H]
     \centering
     \includegraphics[scale=0.7]{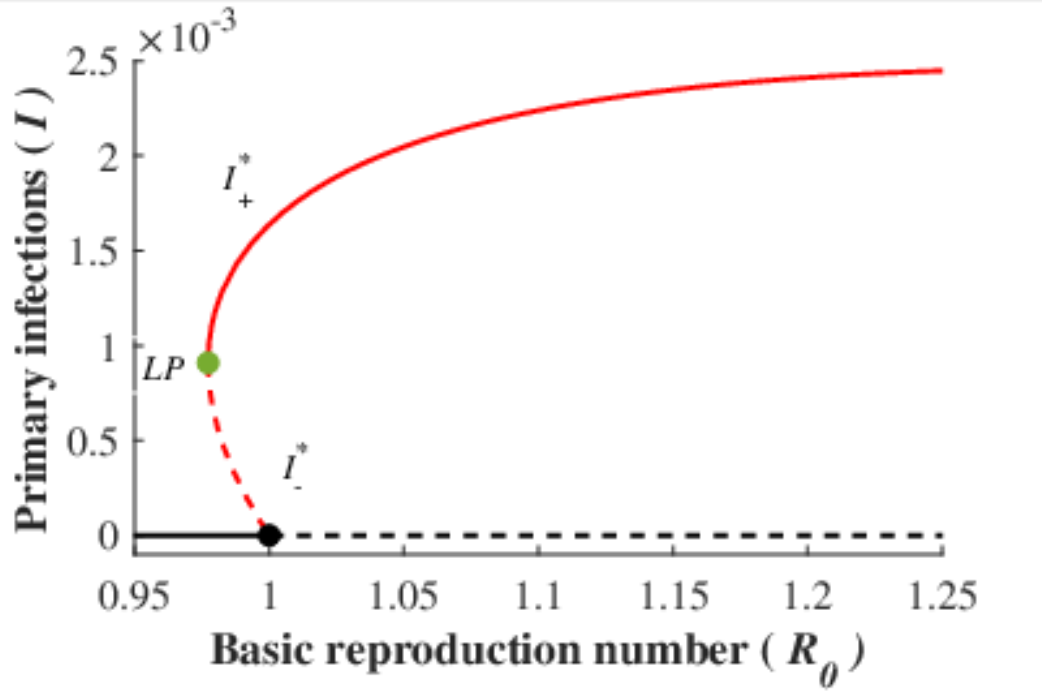}
     \includegraphics[scale=0.7]{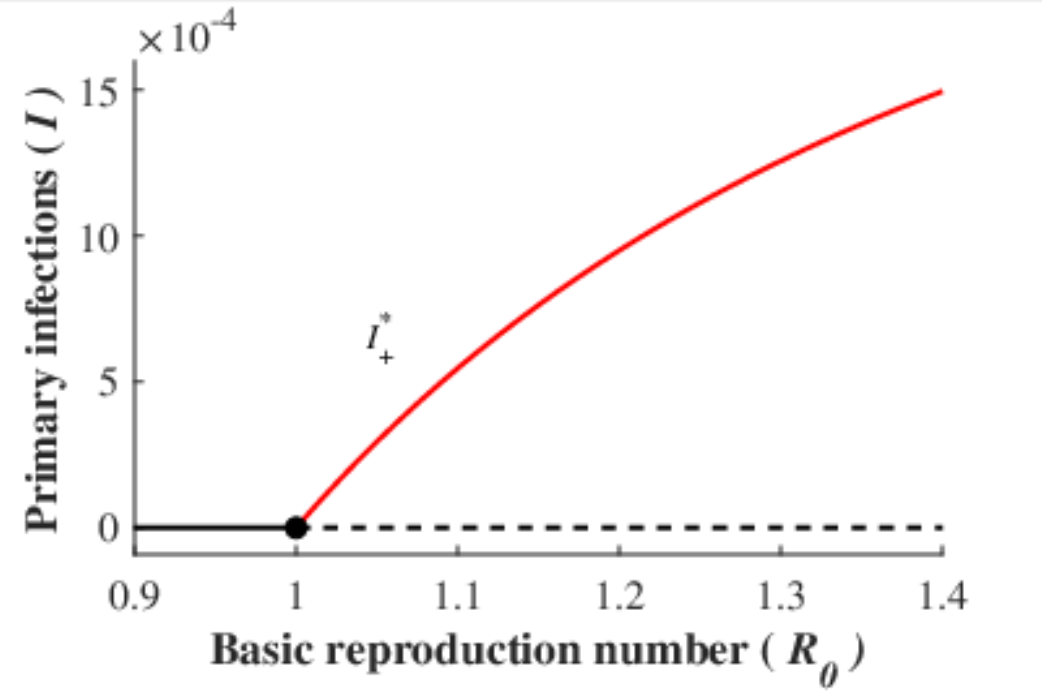}
     \caption{Backward bifurcation (left) and forward bifurcation (right) at $\rzero=1$. Stable branches are marked with continuous and unstable branches with dashed lines. 
     Note that the depicted stability may be lost for $\rzero \gg 1$ as it will be discussed in later sections.
     The parameters used for both cases are $\rho=17, \kappa=0.1, \gamma=17, \mu=0.0125, \nu=3, \alpha=2$. The relative infectivity in the backward case is $\xi=0.9$ and in the forward case $\xi=0.001$.}
     \label{fig:Transcrit-bif}	
\end{figure}

Let us summarize our epidemiologically feasible findings. Depending on the parameters in the system, there can be two types of bifurcations at $\rzero = 1$, forward (supercritical) or backward (subcritical), Figure~\ref{fig:Transcrit-bif}. In a forward bifurcation, a small positive asymptotically stable equilibrium appears and the disease free equilibrium loses its stability at $\rzero=1$. On the other hand, in a backward bifurcation, a branch of unstable endemic equilibria emerges from the DFE.  

This phenomenon was also observed for example in \cite{Heidecke2021, KATRIEL2019}, where the qualitative properties of a simple two-stage contagion model was investigated. The backward bifurcation case is of particular importance as it leads to a bistable situation and the potential persistence of the disease in the population even for $\rzero<1$.  

Moreover, when $\Theta>0$, i.e., the backward bifurcation case, the system undergoes a saddle-node bifurcation at a certain $\tilde\beta\in(0,\gamma+\mu)$ 
the existence of which is established in Theorem~\ref{thm:otherEquilibria}. The saddle-node bifurcation point is marked with LP (limit point) on the equilibria branch. The upper branch of LAS positive equilibria extends beyond $\rzero>1$ which corresponds to the unique EE branch. In Section~\ref{sec:RH}, we will analyze the local stability of the EE for $\rzero \gg 1$ and observe the possibility of 
both loosing and regaining local stability depending on the boosting coefficient, the partitioning of the period of immune protection, and the relative infectivity. 

\subsection{Case \texorpdfstring{$\xi = 0$}{ξ = 0}}
\label{sec:extremeCase}
In the analysis so far, we assumed $\xi > 0$. By considering 
a non-infectious $J$ compartment, i.e. $\xi = 0$, 
the derivation of the formulae is slightly different. 
We omit details of the entire calculation here and 
only share the results. 

The expressions \eqref{eq:Sstar}, \eqref{eq:Istar}, 
\eqref{eq:Jstar}, \eqref{eq:Wstar}, and \eqref{eq:quad_eq_coeff} 
for the equilibria (other than the DFE) remain valid. 
The bound $\Wupper$ becomes infinity indicating that any 
root of the quadratic equation that is conforming the 
lower bound in \eqref{eq:Wbounds} leads to an 
epidemiologically feasible equilibrium. 
Moreover $\SuperCond < 0$, so it is guaranteed that 
a transcritical bifurcation of forward-type occurs at 
$\rzero = 1$ and that no other equilibrium of interest exists 
for $\rzero \leq 1$. For $\rzero > 1$, it is easy to see 
that the lead coefficient of \eqref{eq:Wstar} is negative
\begin{equation*}
    A|_{\xi = 0} = \nu \beta^2 \Big[
    \big( \alpha \kappa (\rho + \mu) - 
     \eta \kappa (\rho + \mu) \big) Q2
    \Big] = - \nu \beta^2 \omega \kappa (\rho + \mu) Q_2 < 0,
\end{equation*}
hence, the parabola $f$ 
is downward with the positive $y$-intercept $C$.
These ensure the existence and uniqueness 
of the endemic equilibrium. 

For an in-depth analysis, the reader is referred to
our computer algebra codes \cite{github}.

\subsection{Stability of the endemic equilibrium for \texorpdfstring{$\rzero > 1$}{R0 > 1}}\label{sec:RH}

The Jacobian evaluated at the endemic equilibrium is 
\begin{equation}
\label{eq:Jacobian_1}
\jac = \left[ 
\begin{array}{cccc}
    - \mu - \beta (\Ieq + \xi \Jeq) & 
        - \beta \Seq & 
        \omega \kappa & - \xi \beta \Seq \\
    \beta (\Ieq + \xi \Jeq) & 
        \beta \Seq - \gamma - \mu & 
        0 & \xi \beta \Seq \\
    - \alpha \kappa & 
        - \alpha \kappa - \nu \beta \Weq & 
        - \alpha \kappa - \mu -\omega\kappa & 
        - \alpha \kappa - \beta \nu \xi \Weq \\
    & & 
        ~ ~ ~ ~ - \nu \beta (\Ieq + \xi \Jeq) & \\
    0 & 
        \nu \beta \Weq & 
        \nu \beta (\Ieq + \xi \Jeq) &
        \beta \nu \xi \Weq - \rho - \mu
  \end{array}\right] 
\end{equation}
yielding the characteristic equation
\begin{equation}
\label{eq:characteristic}
    \det(\jac - \lambda I) = 
    \lambda^4 + a_1 \lambda^3 + a_2 \lambda^2 + 
    a_3 \lambda + a_4 = 0
\end{equation}
with $a_4 = \det(\jac)$.

In order to analyze the stability of the EE, 
we shall use the Routh-Hurwitz criterion \cite{murray2007mathematical,routh1877treatise} that gives information 
on the sign of the real parts of the roots of 
\eqref{eq:characteristic} through inequalities 
formulated in terms of $a_i$. 

\begin{theorem}[Routh-Hurwitz]
\label{thm:RH}
Let $\rzero > 1$, EE as given by \eqref{eq:Sstar}, 
\eqref{eq:Istar}, \eqref{eq:Jstar}, 
\eqref{eq:Wstar-endemic} and $\jac$ the 
Jacobian evaluated there as in \eqref{eq:Jacobian_1}. 

Then, EE is locally asymptotically stable if and only if the 
coefficients of the characteristic polynomial 
\eqref{eq:characteristic} satisfy 
\begin{enumerate}[label=(\roman*)]
    \item $a_i > 0$ for $i = 1, 2, 3, 4$,
    \item $a_1 a_2 > a_3$, \qquad and 
    \item $a_1 a_2 a_3 > a_1^2 a_4 + a_3^2$.
\end{enumerate}
\end{theorem}
First, note that {\it (ii)} can be derived 
from the other two conditions. Then, let's turn our 
attention to the positivity of the coefficients
that is to condition {\it (i)}. 

Using that
\begin{equation*}
\label{eq:IstarPlusXiJstar}
    \Ieq + \xi \Jeq = 
        \frac{\Ieq (\rho + \mu)}{
        \rho + \mu - \beta \nu \xi \Weq}
\end{equation*}
by \eqref{eq:Jstar} and the formula 
\eqref{eq:Sstar}, we obtain
\begin{equation*}
\label{eq:Jacobian}
\jac = \left[ 
\begin{array}{cccc}
    - \mu - \frac{\Ieq \beta (\rho + \mu)}{
         \rho + \mu - \beta \nu \xi \Weq} & 
        - \frac{(\gamma +\mu) 
            (\rho + \mu -
             \beta \nu \xi \Weq)}{\rho + \mu} & 
        \omega \kappa & - \frac{\xi (\gamma +\mu ) 
            (\rho + \mu - \beta \nu \xi \Weq)}{
                \rho + \mu} \\
    \frac{\Ieq \beta (\rho + \mu)}{
         \rho + \mu - \beta \nu \xi \Weq} & 
        \frac{\beta \nu \xi \Weq (\gamma + \mu)}{
            \rho + \mu} & 
        0 & \frac{\xi (\gamma +\mu ) 
            (\rho + \mu - \beta \nu \xi \Weq)}{
                \rho + \mu} \\
    - \alpha \kappa & 
        - \alpha \kappa - \nu \beta \Weq & 
        - \alpha \kappa - \mu -\omega\kappa & 
        - \alpha \kappa - \beta \nu \xi \Weq \\
    & & 
        ~ ~ ~ ~ - \frac{\Ieq \nu \beta (\rho + \mu)}{
         \rho + \mu - \beta \nu \xi \Weq} & \\
    0 & 
        \nu \beta \Weq & 
        \frac{\Ieq \nu \beta (\rho + \mu)}{
         \rho + \mu - \beta \nu \xi \Weq} &
        - \rho - \mu + \beta \nu \xi \Weq 
  \end{array}\right].
\end{equation*}
When expanding  
$\det(\jac - \lambda I)$, terms appear with 
positive and negative signs in each expression. 
We employed a series of operations  
grouping \emph{all} negative ones with \emph{some} of the 
positive terms leading to simplified residual expressions. 
For the technical details, we refer to the supplementary 
computer algebra codes \cite{github}. 
As not all positive terms were used, these residuals 
may serve as lower bounds on $a_i$ and are listed below 
\begin{subequations}
\label{eq:coefficients}
\begin{align}
    a_1 &= 
        \eta \kappa + 2 \mu + 
        \frac{\gamma + \mu}{\rho +\mu} \beta \nu \xi \Weq + 
        \Big( \rho + \mu - \beta \nu \xi \Weq \Big) + 
        \frac{\beta \Ieq (\nu + 1) (\rho + \mu)}{
            \rho + \mu - \beta \nu \xi \Weq},
    \label{eq:a1} \\
    a_2 &> 
    \beta \Ieq (\rho + \mu) + \rho \frac{\eta \kappa + 2 \mu}{\rho + \mu} \Big( \rho + \mu - \beta \nu \xi \Weq \Big), 
    \label{eq:a2} \\
    a_3 &> 
    \beta \Ieq (\rho + \mu) 
    \Big( \eta \kappa 
     + \gamma + 2 \mu \Big) +
     \Big( \rho + \mu - \beta \nu \xi \Weq \Big)
     \rho \frac{\eta \kappa (\kappa + \mu) + \mu^2}{\rho + \mu}, 
     \label{eq:a3} \\
    a_4 &> 
    - \beta \Ieq \SuperCond.
    \label{eq:a4}
\end{align}
\end{subequations}
Clearly, the positivity of $a_i$ for $i = 1, 2, 3$ is 
established by \eqref{eq:a1}, \eqref{eq:a2}, \eqref{eq:a3} 
as \\ $\rho + \mu - \beta \nu \xi \Weq > 0$ must hold 
by \eqref{eq:Jstar} and by 
the positivity of the components of the EE. In addition, 
we see that assuming $\SuperCond \leq 0$ ({\it i.e.} the case of
forward transcritical bifurcation) 
readily implies the 
positivity of $a_4$ in \eqref{eq:a4}. 

To fully analyze this final coefficient, let us 
recall that $a_4 = \det(\jac)$. In order 
to obtain an alternative bound, we carry out 
a series of transformations on $\jac$ in 
\eqref{eq:Jacobian_1}, all of which 
are preserving the sign of the determinant with 
the intermediate goal of obtaining a tractable 
row-echelon form. These 
transformations fall into four categories:
\begin{enumerate}
    \item multiplication from left or right by a 
    matrix with positive determinant:
    \begin{itemize}
        \item[-] scaling of a row/column 
        by a positive number; 
        \item[-] multiple row and column changes 
        given by permutation matrices with 
        $\det = 1$; 
        \item[-] carrying out row/column 
        elimination towards the echelon form; 
    \end{itemize}
    \item adding the zero matrix:
    \begin{itemize}
        \item[-] use \eqref{eq:sirw_reduced_equlibria} 
        to hop back-and-forth between 
        transmissional and transitional terms; 
    \end{itemize}
    \item substitution of \eqref{eq:Sstar}, 
        \eqref{eq:Istar}, and \eqref{eq:Jstar}; 
    \item algebraic manipulation of expressions.
\end{enumerate}
Again, the exact steps of this procedure are documented in 
the supplementary computer algebra codes \cite{github}. Here, we just present the final 
form obtained from the reduction 
that is the matrix $\tilde{\jac}$ 
such that $\sgn(a_4) = \sgn(\det(\jac)) = \sgn(\det(\tilde{\jac}))$: 
\begin{equation*}
\label{eq:sgnDetJ}
\tilde{\jac} = 
\left[
\begin{array}{cccc}
 1 
    & 0 
    & 0 
    & \beta \nu \xi 
        \Big( \frac{\mu}{\rho + \mu} Q_1 + 
        \eta \kappa^2 \gamma\Big) \\
 0 & 
    1 & 
    - \frac{1}{F_1} & 
    - \left(
        F_1 - \frac{\mu (\beta - \gamma - \mu)}{\beta \Weq}
      \right) Q_0 \\
 0 & 
    0 & 
        1 & 
    F_2 \Big( \frac{\mu}{\rho + \mu} Q_1 + 
        \eta \kappa^2 \gamma\Big) \\
 0 & 
    0 & 
    0 & 
    F_1^2 Q_0 - 
    F_2 \Big( \frac{\mu}{\rho + \mu} Q_1 + 
        \eta \kappa^2 \gamma\Big)\\
\end{array}
\right],
\end{equation*}
where 
\begin{equation*}
\label{eq:Fi}
    F_1 =  
        \left((\gamma + \mu) + 
        (\beta - \gamma - \mu)
            \frac{\rho + \mu}{\beta \nu \xi \Weq} \right)
             \frac{\mu \nu \xi}{\mu + \rho} + 
             \omega \kappa
    \qquad \text{and} \qquad 
    F_2 = \beta \frac{\mu \nu \xi}{\mu + \rho} + 
        \omega \kappa .
\end{equation*}
Clearly $F_1 > F_2 > 0$ by $\rzero > 1$, 
\eqref{eq:Jstar} and $\Ieq, \Jeq, \Weq > 0$, 
hence, it suffices to show that 
\begin{equation*}
  F_2 Q_0 - 
    \Big( \frac{\mu}{\rho + \mu} Q_1 + 
        \eta \kappa^2 \gamma\Big)
\end{equation*}
is positive.
Note that 
\begin{equation*}
    F_2 Q_0 = 
    \frac{\mu}{\rho + \mu} \beta \nu \xi Q_0 + 
        \eta \kappa^2 \gamma
\end{equation*}
leads to analyzing the sign of 
\begin{equation*}
    \beta \nu \xi Q_0 - Q_1.
\end{equation*}
Then, as 
$\beta \nu \xi Q_0 - Q_1 > \SuperCond$ when $\rzero > 1$, 
we obtain the positivity of $a_4$ for $\SuperCond > 0$.
Hence, using the implications of \eqref{eq:a4} 
when $\SuperCond \leq 0$, we established  
that $a_4 > 0$ is satisfied that is  
condition {\it (i)} 
of Theorem~\ref{thm:RH} holds.

Therefore, by defining 
\begin{equation}
\label{eq:RH}
  y_{\nu}(\alpha, \xi) = 
  a_1 a_2 a_3 - (a_1^2 a_4 + a_3^2), 
\end{equation}
all conditions of 
Theorem~\ref{thm:RH} are satisfied if and only if
$y_{\nu}(\alpha, \xi) > 0$. When $\xi$ is fixed, 
we use the notation $y_\nu(\alpha)$. 
In the following, condition \eqref{eq:RH} is referred 
to as the Routh-Hurwitz criterion. The sign of 
\eqref{eq:RH} will be studied using numerical 
techniques in the next section.

\section{Exploring bifurcations using numerics}
\label{sec:numericalExploration}
In this section, we investigate numerically how the asymmetric partition of the immunity period, the boosting rate, and the relative infectivity influence the stability changes of the EE. Of particular interest are the formation of bistability regions influenced by the relative infectivity $\xi$. 

For our numerical investigations, we set the parameters as
\begin{equation}
\label{eq:parametrization}
\begin{split}
    &\gamma = 17, \\
    &\kappa = 1/10, \\
    &\mu = 1/80, \\
    &\beta=260, \\
    &\rho =17, \\
\end{split}
\end{equation}
and $\xi\in(0,1),$ taken from \cite{lavine2011natural,strube}, where authors studied natural immune boosting in pertussis dynamics. 

\begin{figure}[H]
\centering
\begin{subfigure}{.54\textwidth}
    \centering
    \includegraphics[scale=0.78]{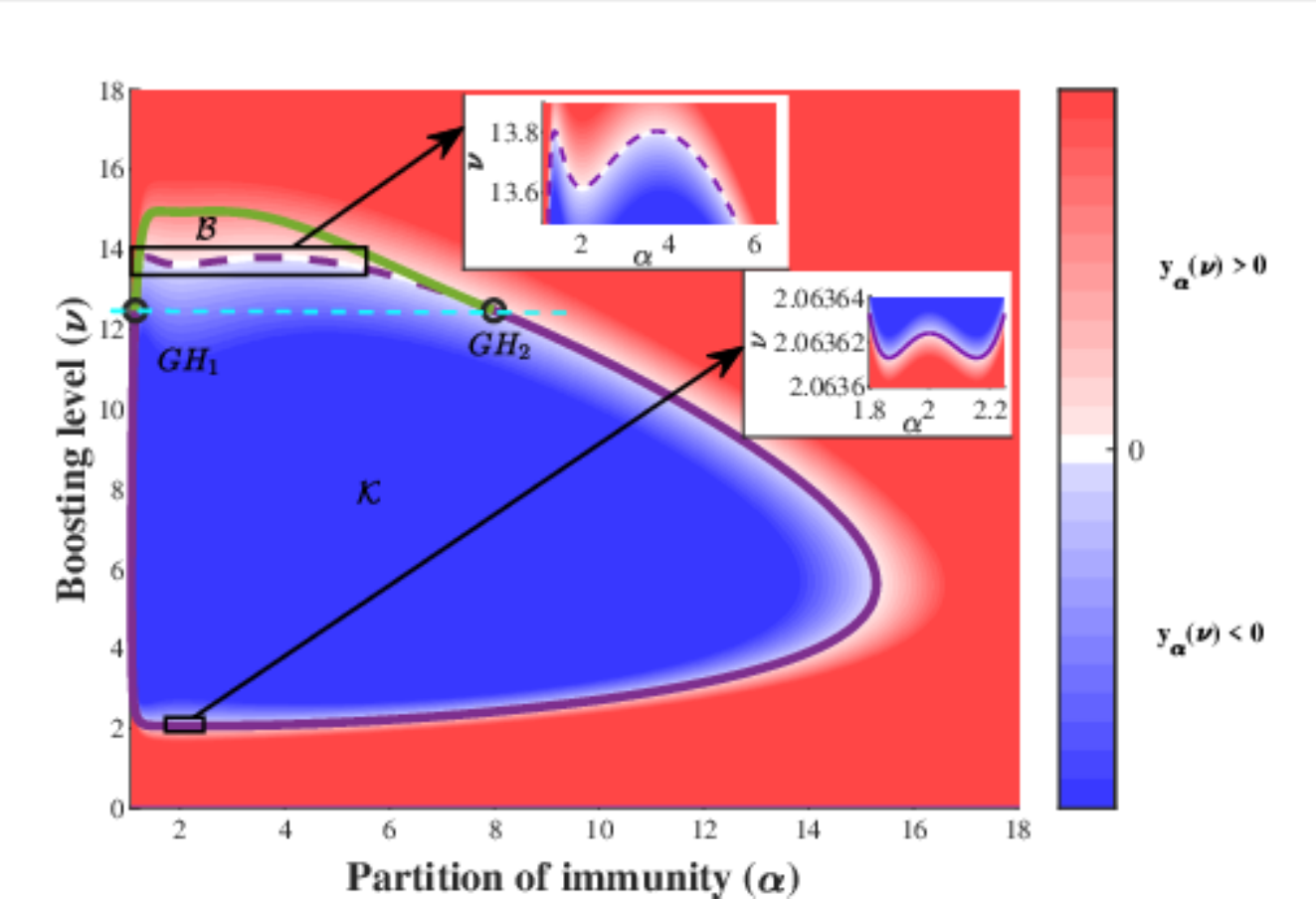}
    \captionsetup{
        margin = 20pt
    }
    \caption{}
    \label{fig:RH-Large-Left2}
\end{subfigure}~\qquad
\begin{subfigure}{.4\textwidth}
    \centering
    \includegraphics[scale=0.75]{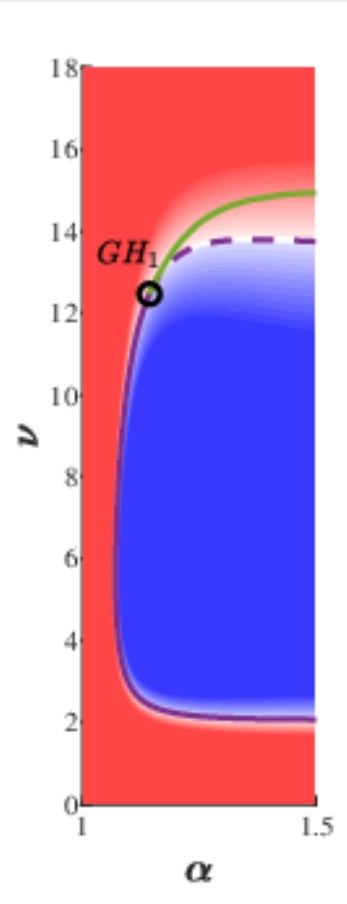}
    \captionsetup{
        margin = 20pt
    }
    \caption{}
    \label{fig:RH-Large-Right2}
\end{subfigure}
    \captionsetup{
        margin = 0pt
    }
    \caption{Baseline dynamics: $\xi = 0$, $\rho \to \infty$. Heatmap of the Routh-Hurwitz stability criterion and bistability region. Purple curve represents $y_\nu(\alpha) = 0$.}
    \label{fig:heatmap-previous}
\end{figure}

In a former work \cite{Richmond1}, 
a similar epidemic model (SIRWS) 
was investigated 
where the $J$ compartment was absent and 
boosting resulted in immediate immunity, 
namely, a return to $R$ from $W$. The current 
system reconstructs the same dynamics 
in the limit that is for $\xi = 0$ and 
$\rho \to \infty$. As a starting point, 
we briefly review the structure of 
the aforementioned scenario via 
Figure~\ref{fig:heatmap-previous}.

\begin{figure}[H]
	\centering
	\begin{subfigure}{0.45\textwidth}
		\includegraphics[scale=0.7]{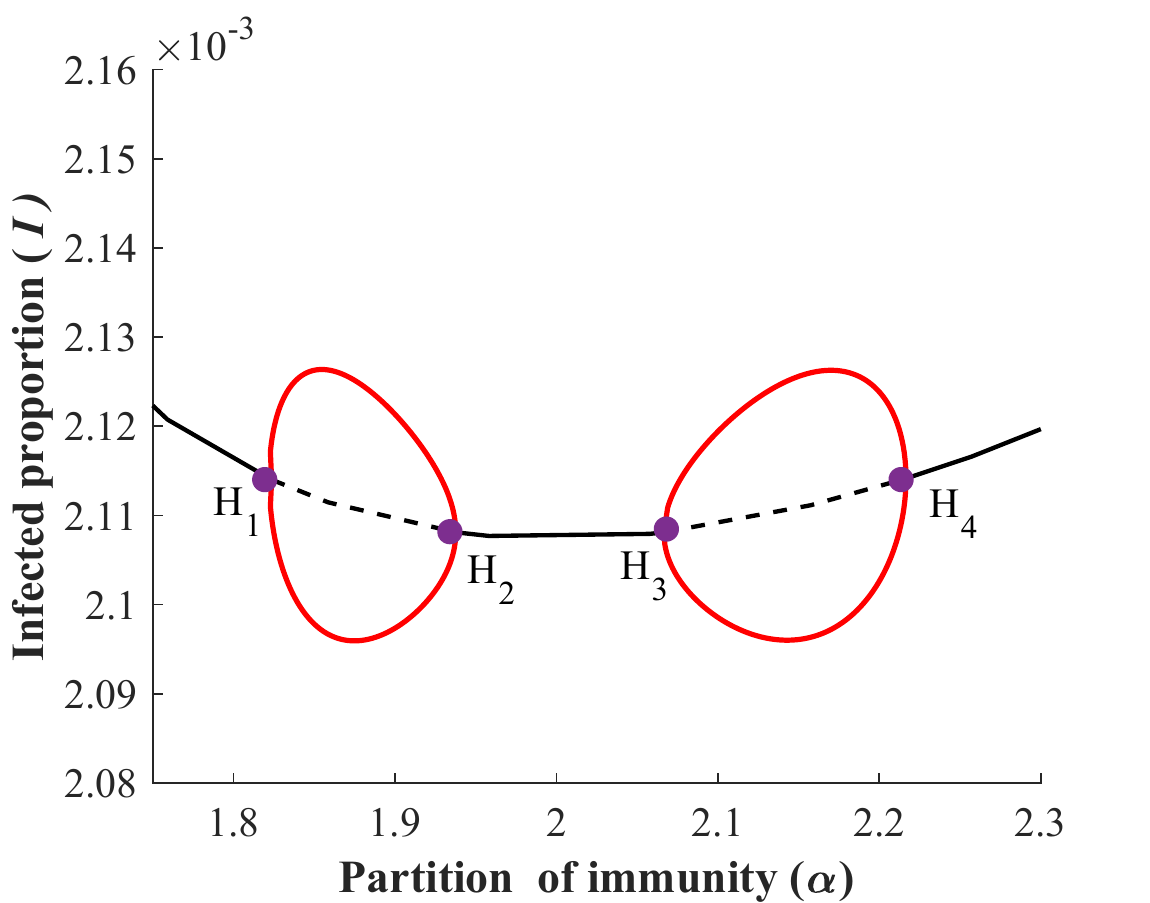}
	\end{subfigure}\qquad
	\begin{subfigure}{0.45\textwidth}
		\includegraphics[scale=0.7]{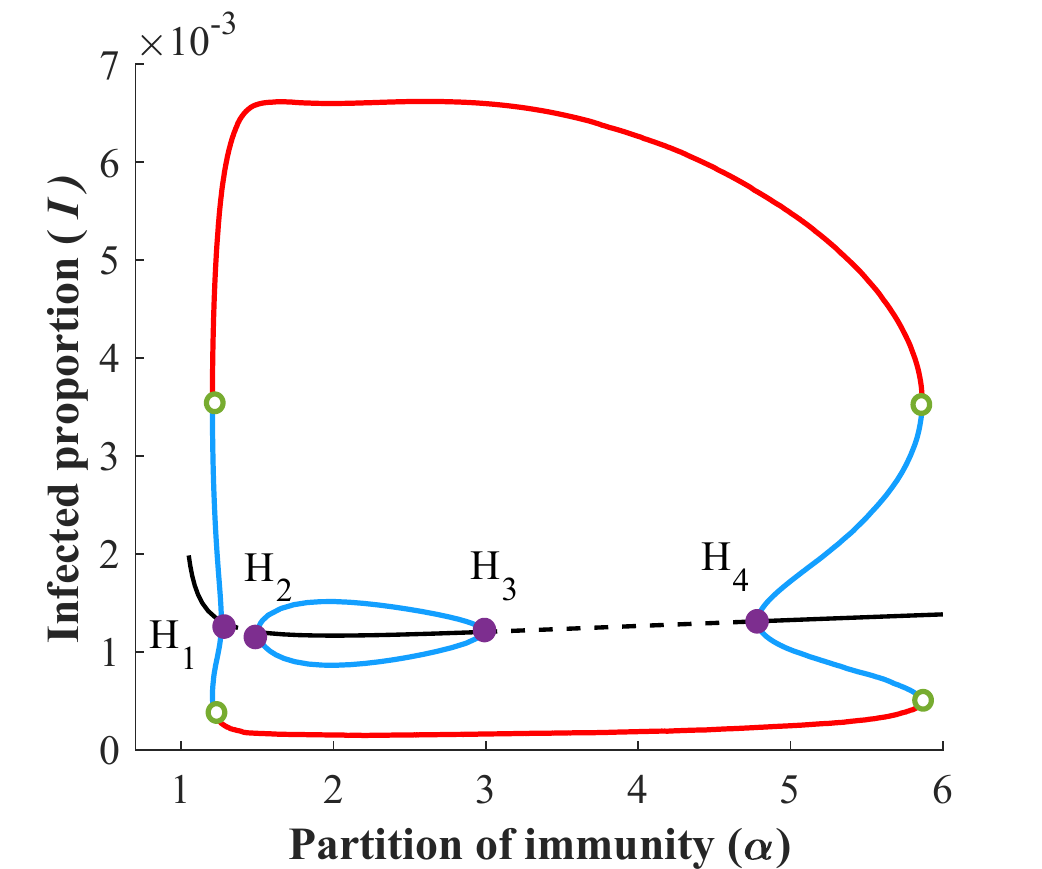}
	\end{subfigure}
	\caption{Baseline dynamics: $\xi = 0$, $\rho \to \infty$. Bifurcation diagram w.r.t $\alpha$, with $\nu = 2.06362$ (left) and  $\nu=13.7$ (right). The depicted two bubbles of instability appear and disappear simultaneously. The later phenomenon is referred to as \emph{symmetric presence of bubbles}.} \label{fig:bif_double_bubble}
\end{figure}

First, we recall that at $\rzero = 1$ the transcritical 
bifurcation was shown to be solely of forward type. At the baseline 
parametrization $\rzero \approx 15.28$ and 
the endemic equilibrium is LAS but for the 
compact set $\mathcal{K}$ marked by blue. 
Note the symmetric presence of epidemic 
double bubbles around the 
baseline partitioning $\alpha = \omega = 2$ in two boosting ranges, as highlighted in the insets of Figure~\ref{fig:RH-Large-Left2}. 
We recall two examples from \cite{Richmond1}, where the boosting values $\nu \approx 2.06362$ 
and $\nu \approx 13.7$ in these regions lead to the appearance of double bubbles, see Figure~\ref{fig:bif_double_bubble}.     

For slightly 
larger boosting $\nu \approx 14$, a bistable 
region $\mathcal{B}$ was observed where the 
EE is LAS 
together with a stable periodic orbit.
The appearance of this bistable region is characterized 
by two \emph{generalized Hopf points} $GH_1$ and $GH_2$ with identical 
$\nu$ coordinates.

Now, focusing on the current model and  
parametrization, 
first, we briefly study the direction of the transcritical
bifurcation that is determined by the sign of $\SuperCond$ in 
Section~\ref{sec:numericalSupercond}, second, 
we analyze the stability of the EE 
through sign analysis of the Routh-Hurwitz criterion in 
Section~\ref{sec:EE-stability}. Then, 
we carry out numerical analysis of the 
bifurcations of the equilibrium branch and study 
how the bistable region is affected by the relative 
infectivity $\xi$ in Section~\ref{sec:matcont}.

\subsection{Direction of the transcritical bifurcation}
\label{sec:numericalSupercond}

\begin{figure}[H]
    \centering
    \includegraphics[width=0.7\textwidth]{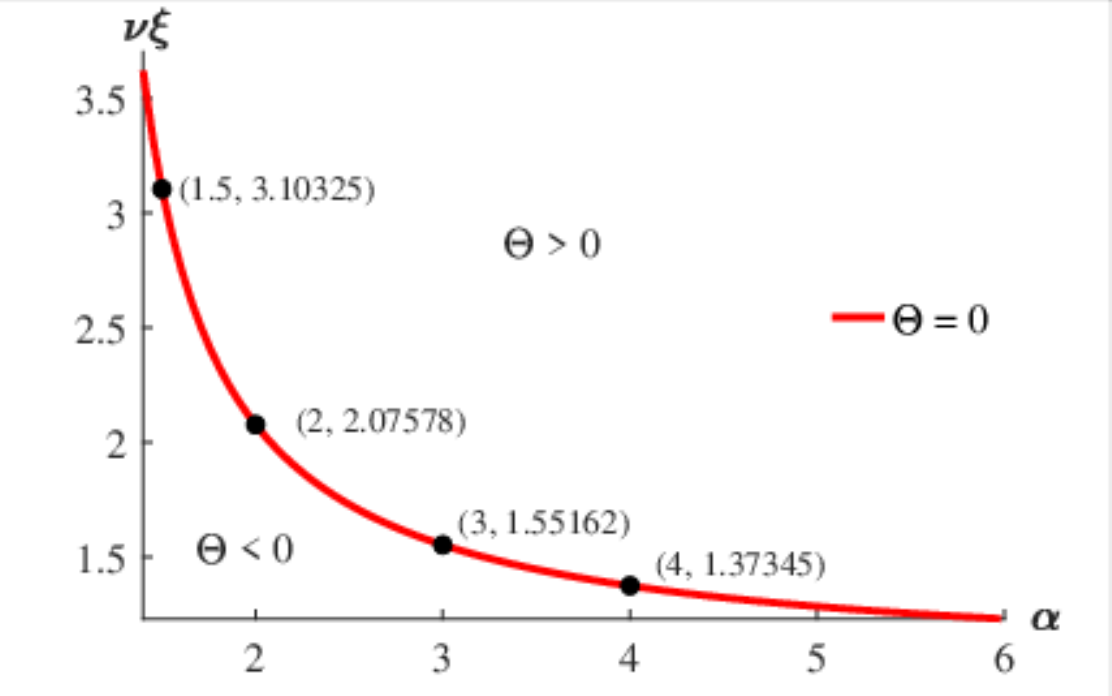}
    \caption{$\Theta = 0$ contour on the $(\alpha, \nu \xi)$ plain using 
    the parametrization \eqref{eq:parametrization}.}
    \label{fig:SuperCond}
\end{figure}

Substituting the baseline parametrization \eqref{eq:parametrization} into 
\eqref{eq:SuperCond}, we have that $\SuperCond > 0$ is equivalent to
\begin{equation*}
    \nu \xi > \frac{1.00662 \alpha}{\alpha - 1} + \frac{0.125092}{\alpha}
    =: \SuperBound(\alpha)
\end{equation*}
with the corresponding zero contour displayed 
in Figure~\ref{fig:SuperCond}. 

Clearly $\lim_{\alpha \to 1^+}\SuperBound(\alpha) = \infty$ and 
$\lim_{\alpha \to \infty}\SuperBound(\alpha) = 1.0062$, 
moreover, $\SuperBound(\alpha)$ is decreasing 
function of $\alpha$. 
Consequently, the faster the transition from 
$W$ to $R$, the smaller boosting coefficient $\nu$ is sufficient to 
activate backward transcritical bifurcation at $\rzero = 1$ that is 
at $\beta = \gamma + \mu = 17 + 1/80$ 
while keeping the relative infectivity $\xi$ fixed, or 
vice versa, smaller $\xi$ is required with keeping $\nu$ fixed. 
For example, assuming a moderate boosting coefficient,  
{\it i.e.} $\nu < 3$, and a relative infectivity $\xi \sim O(1)$ 
may very well result in a backward bifurcation for $\alpha \geq 1.5$.

\subsection{Stability switches of the EE}
\label{sec:EE-stability}
\begin{figure}[H]
\centering
\begin{subfigure}{.54\textwidth}
    \centering
    \includegraphics[scale=0.78]{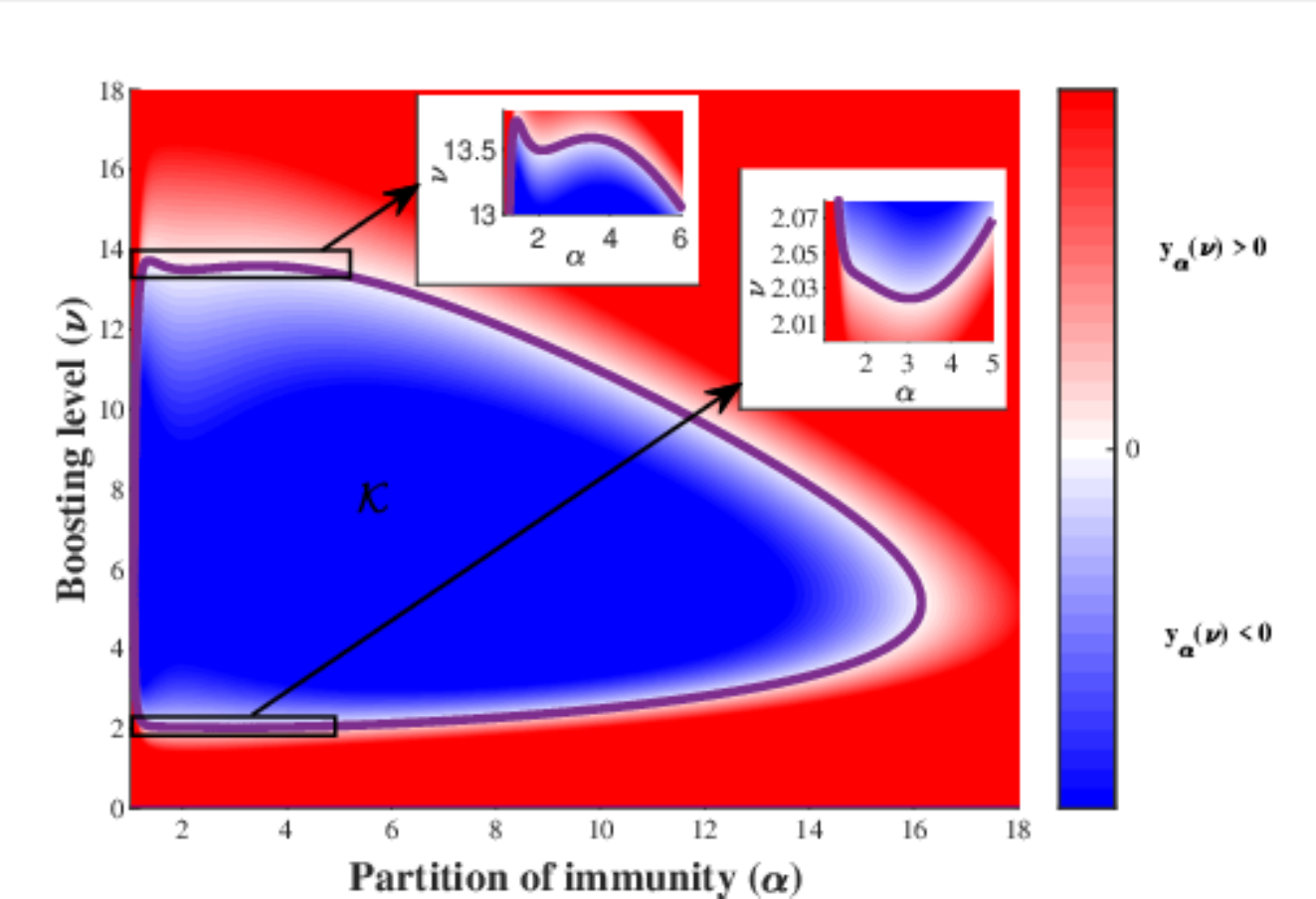}
    \captionsetup{
        margin = 20pt
    }
    \caption{}
    \label{fig:RH-Large-Left}
\end{subfigure}~\qquad
\begin{subfigure}{.4\textwidth}
    \centering
    \includegraphics[scale=0.75]{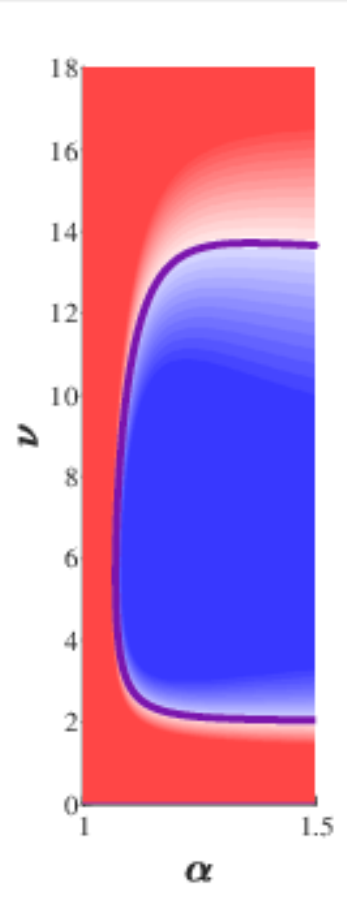}
    \captionsetup{
        margin = 20pt
    }
    \caption{}
    \label{fig:RH-Large-Right}
\end{subfigure}
    \captionsetup{
        margin = 0pt
    }
    \caption{Heatmap of the Routh-Hurwitz criterion \eqref{eq:RH} for $\xi = 0$. Purple curve represents $y_{\nu}(\alpha)=0$.}
    \label{fig:Heatmap-zero}
\end{figure}

We constructed similar heatmaps to study the sign of 
$y_\alpha(\nu,\xi)$, given by \eqref{eq:RH}, 
for various values of relative infectivity $\xi \geq 0$.
Recall that $\rho = \gamma = 17$ in 
our setting, thus, for $\xi = 0$ 
we readily experience changes in the 
dynamics with respect to 
Figure~\ref{fig:heatmap-previous}. 
The instability set, marked as $\mathcal{K}_\xi$ to emphasize its dependence of $\xi$, 
is somewhat similar but 
the regular shape resulting 
in simultaneous appearance of double-bubbles 
of instability is lost, 
see Figure~\ref{fig:Heatmap-zero}. Note that in all figures that follow, $\mathcal{K}_\xi=\mathcal{K}$ for fixed $\xi$. Now, the region around 
$\nu \approx 2.06$ displays much simpler 
behaviour. Additionally, 
for $\nu \approx 13.5$,
we still see  
bubbles, though without the symmetry 
they possess in the limit $\rho \to \infty$. 

By increasing $\xi\in (0,1)$, 
we observe the following two phenomena. 
First, the shape of the 
$y_\alpha(\nu,\xi) =0$ curve that bounds 
the set $\mathcal{K}_\xi$ changes, therefore it 
influences the number of stability switches of
the EE in the $(\alpha,\nu)$ plane. Second, 
the region $\mathcal{K}_\xi$ is 
shrinking and then disappearing, 
hence it results in the increase of 
local asymptotic stability region of the EE. 

\begin{figure}[H]
\centering
\begin{subfigure}{.54\textwidth}
    \centering
    \includegraphics[scale=0.78]{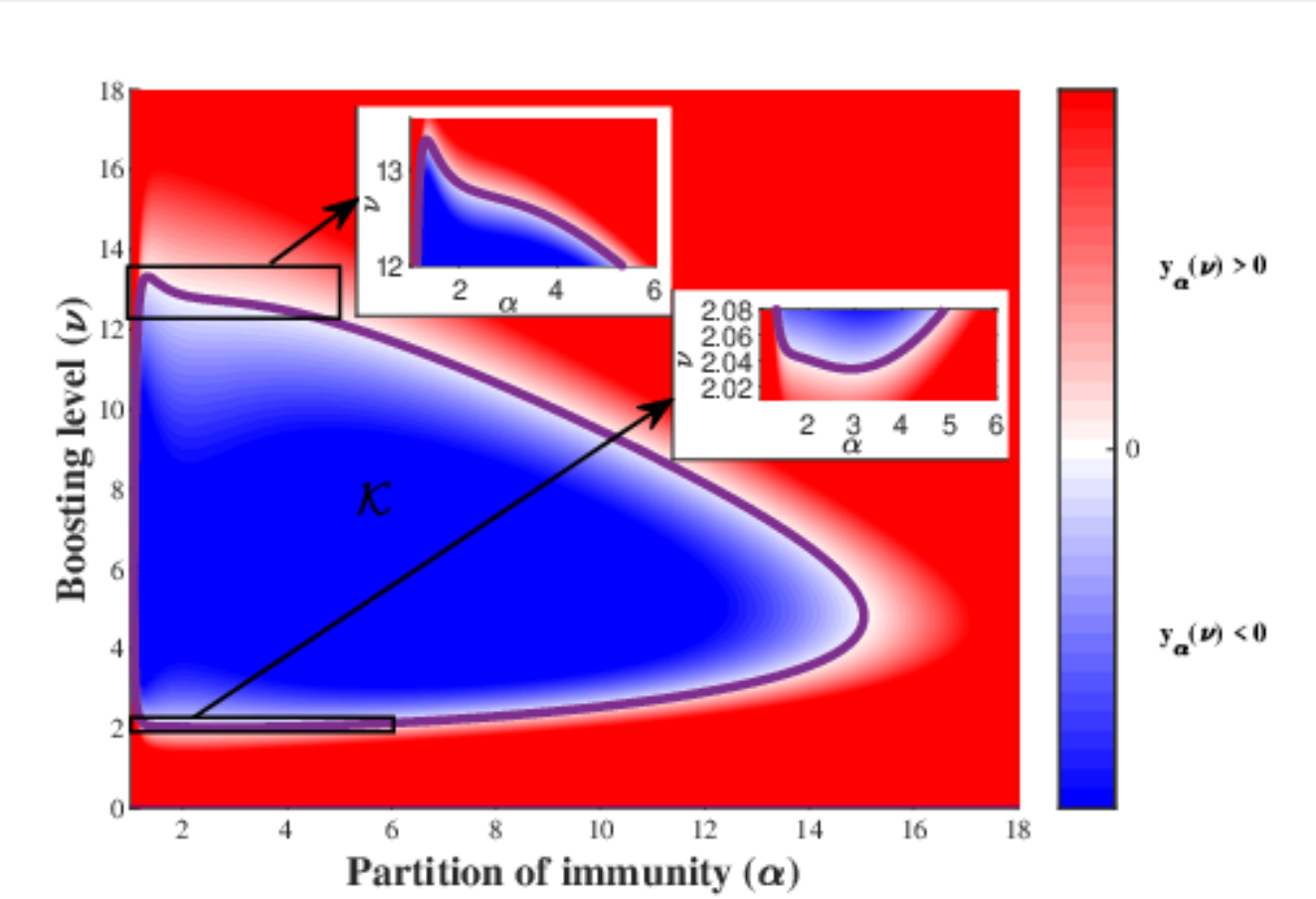}
    \captionsetup{
        margin = 20pt
    }
    \caption{}
    \label{fig:RH-Large-Left1}
\end{subfigure}~\qquad
\begin{subfigure}{.4\textwidth}
    \centering
    \includegraphics[scale=0.75]{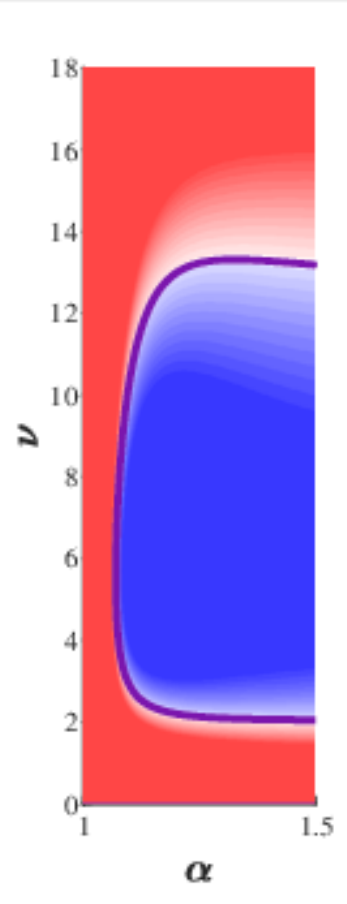}
    \captionsetup{
        margin = 20pt
    }
    \caption{}
    \label{fig:RH-Large-Right1}
\end{subfigure}
    \captionsetup{
        margin = 0pt
    }
    \caption{Heatmap of the Routh-Hurwitz criterion \eqref{eq:RH} for $\xi=10^{-4}$. Purple curve represents $y_{\nu}(\alpha)=0$.}
    \label{fig:Heatmap-e_04}
\end{figure}

\paragraph{Dynamics of stability switches}
For small $\xi$, the Routh-Hurwitz criterion
changes sign multiple times for boosting
rates around $13.5$ as $\alpha$ is varied, 
suggesting the continued 
presence of multiple stability switches 
that is the aforementioned bubbles, see again  
Figures~\ref{fig:Heatmap-zero}.
As $\xi$ grows, the curve 
$y_\nu(\alpha,\xi)=0$ is deforming so that 
these double-bubbles disappear, as in 
Figure~\ref{fig:Heatmap-e_04}. 

\begin{figure}[H]
	\centering
		\includegraphics[scale=0.9]{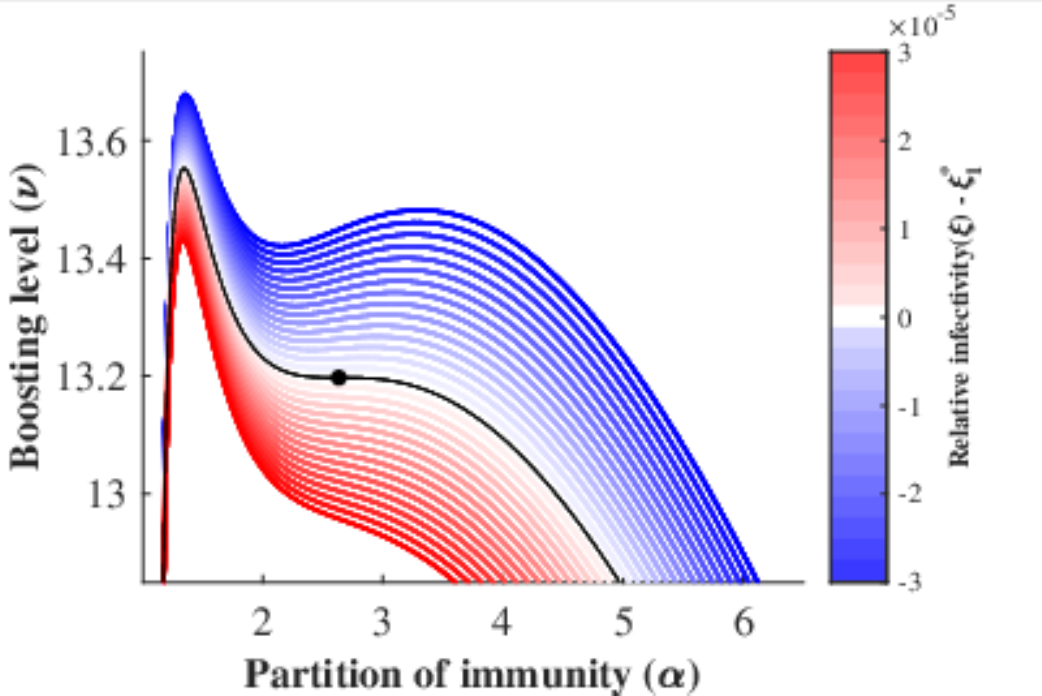}
		\caption{Level curves $y_{\nu}(\alpha,\xi)=0$ for $\xi\in [\xi^*_1 - 3 \times 10^{-5}, \xi^*_1 + 3 \times 10^{-5}]$. The black curve corresponds to the threshold value $\xi^*_{1}$.}
		\label{fig:Stackheat1}
\end{figure}

We localize the 
threshold value $\xi_1^*$, at which 
the relevant change in the qualitative 
behavior of the curve 
$y_\nu(\alpha,\xi)=0$ occurs, 
as follows. In the region of interest 
($1<\alpha < 6$ and $13 < \nu$), 
the level curve 
$\alpha \mapsto \nu ~ \colon y_\nu(\alpha) = 0$,
originally (when $\xi=0$), 
has two local maxima and one local minimum. 
As $\xi$ gets larger, the right maximum and the 
minimum collide, then disappear.
Thus, the threshold scenario may be found by 
looking for $(\nu, \alpha, \xi)$ such that 
\begin{equation*}
\left[
\def\arraystretch{1.3}
\begin{array}{c}
    ~ ~ y_\nu(\alpha,\xi)  \\
    \frac{\partial}{\partial \alpha}y_\nu(\alpha,\xi) \\
    \frac{\partial^2}{\partial \alpha^2} y_\nu(\alpha,\xi)\\
\end{array}
\right] = 
\left[
\def\arraystretch{1.3}
\begin{array}{c}
    0\\
    0\\
    0\\
\end{array}
\right],
\end{equation*}

yielding 
$\xi^*_1\approx 4.0098\times10^{-5}$. 
Figure~\ref{fig:Stackheat1} visualizes 
the transition in the qualitative behavior 
of the curve $y_{\nu}(\alpha,\xi)=0$ 
highlighting the one corresponding 
to the threshold value in black.

\begin{figure}[htb]
    \centering 
\begin{subfigure}{0.25\textwidth}
  \includegraphics[width=\linewidth]{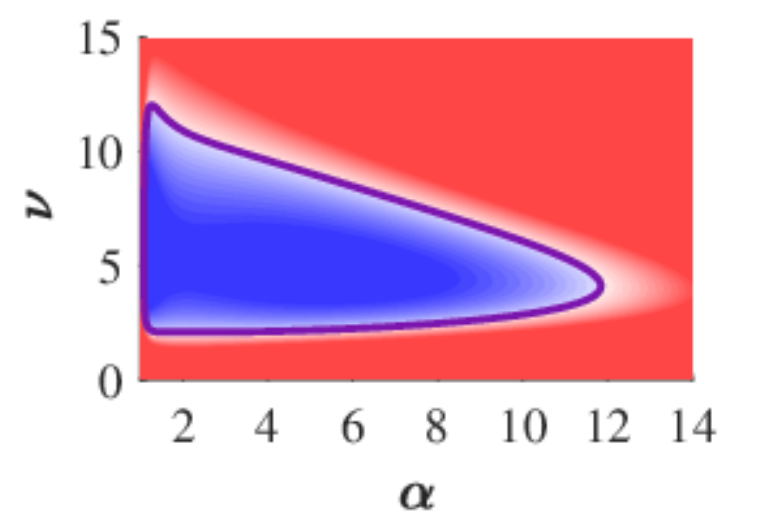}
  \caption{$\xi=5\times 10^{-4}$}
  \label{fig:1}
\end{subfigure}\hfil 
\begin{subfigure}{0.25\textwidth}
  \includegraphics[width=\linewidth]{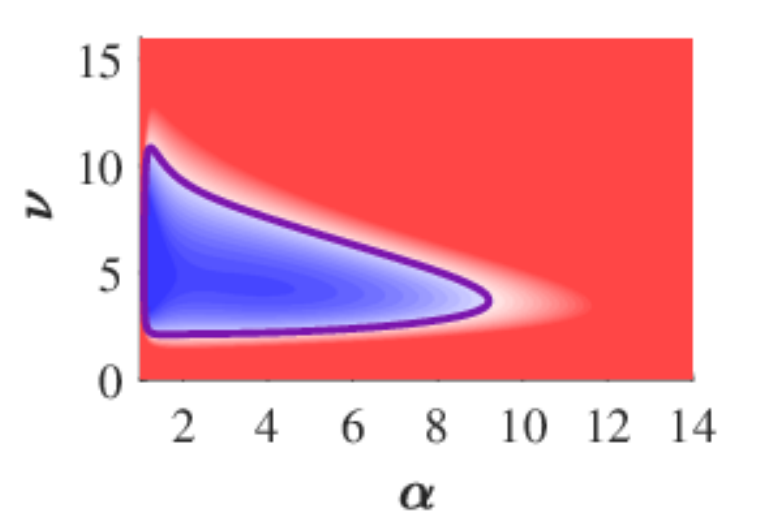}
  \caption{$\xi=1\times 10^{-3}$}
  \label{fig:2}
\end{subfigure}\hfil 
\begin{subfigure}{0.25\textwidth}
  \includegraphics[width=\linewidth]{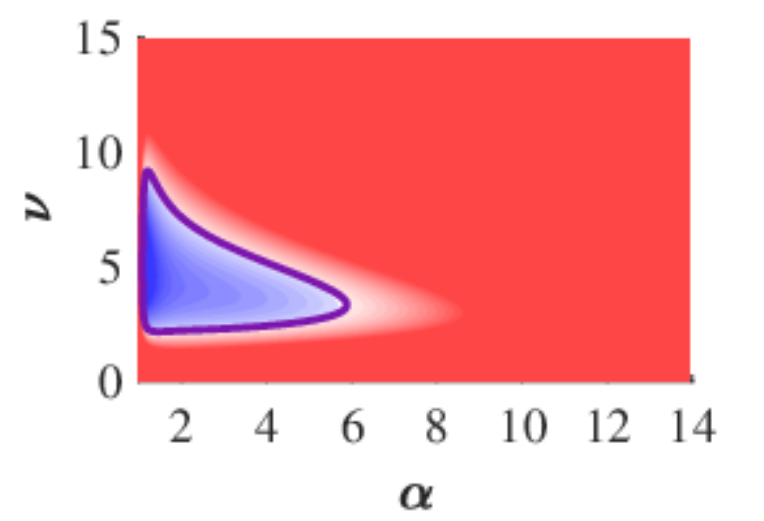}
  \caption{$\xi=2\times 10^{-3}$}
  \label{fig:3}
\end{subfigure}

\medskip
\begin{subfigure}{0.25\textwidth}
  \includegraphics[width=\linewidth]{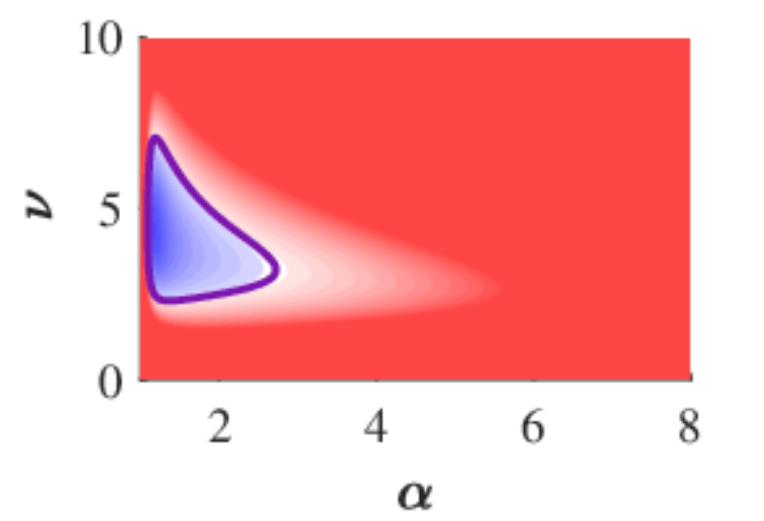}
  \caption{$\xi=4\times 10^{-3}$}
  \label{fig:4}
\end{subfigure}\hfil 
\begin{subfigure}{0.25\textwidth}
  \includegraphics[width=\linewidth]{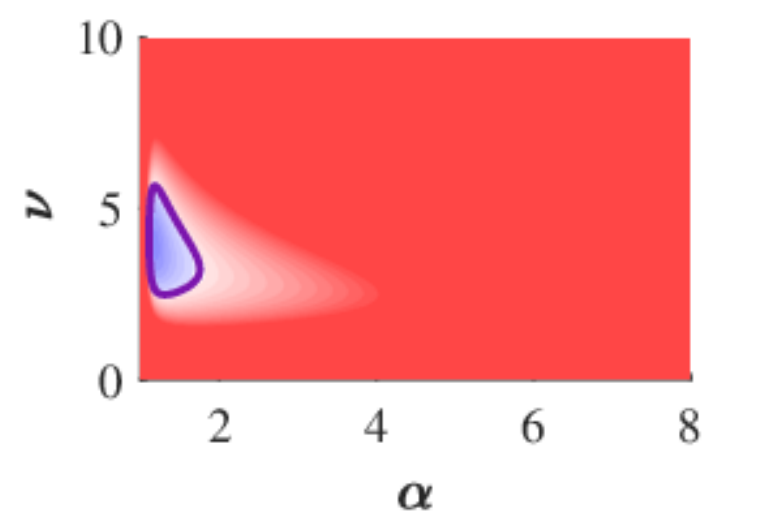}
  \caption{$\xi=6\times 10^{-3}$}
  \label{fig:5}
\end{subfigure}\hfil 
\begin{subfigure}{0.25\textwidth}
  \includegraphics[width=\linewidth]{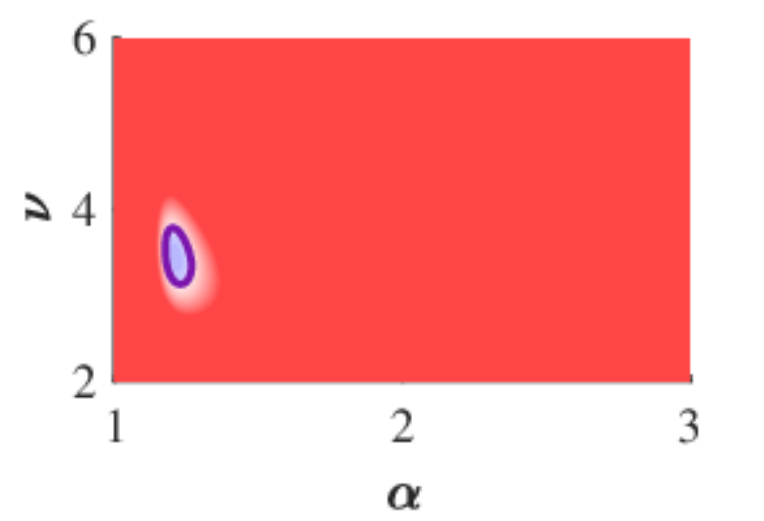}
  \caption{$\xi=9\times 10^{-3}$}
  \label{fig:6}
\end{subfigure}
\caption{Heatmap of the Routh-Hurwitz criterion \eqref{eq:RH}}
\label{fig:Shrinks}
\end{figure}

\paragraph{Shrinking of \texorpdfstring{$\mathcal{K}_\xi$}{K}}
The second phenomenon we analyze is how the 
compact region of instability $\mathcal{K}_\xi$ shrinks and 
disappears as we increase $\xi$, 
see Figure~\ref{fig:Shrinks}.

\begin{figure}[H]
	\centering
		\includegraphics[scale=0.9]{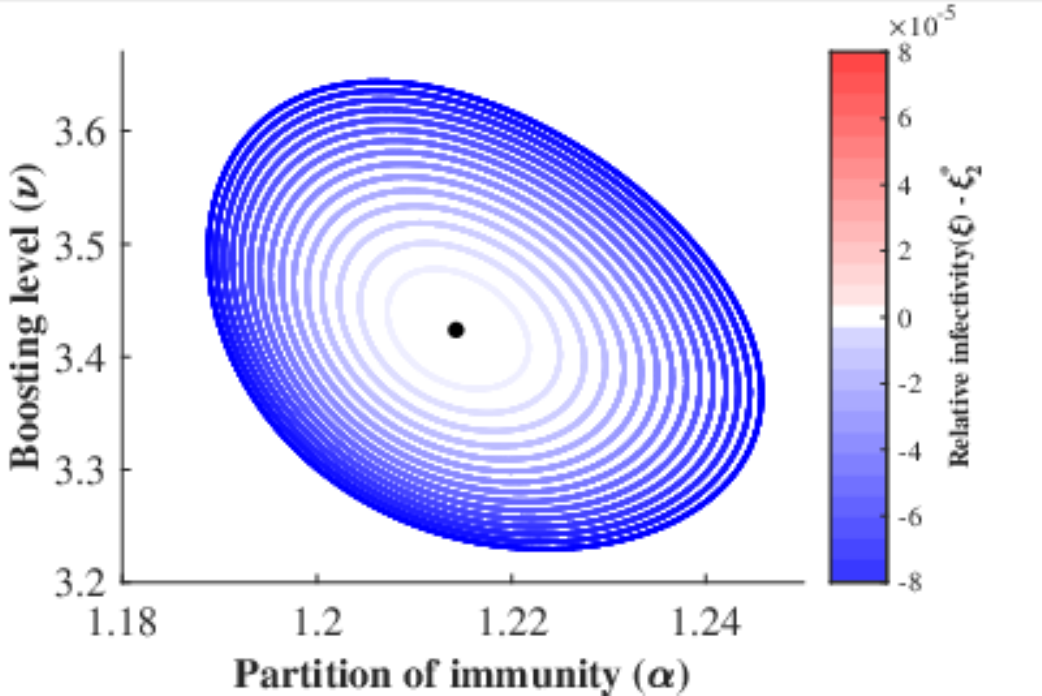}
		\caption{Level curves $y_{\nu}(\alpha,\xi)=0$ for $\xi\in [\xi^*_2 - 8 \times 10^{-5}, \xi^*_2 + 8 \times 10^{-5}]$. The curves cease to exist for $\xi > \xi^*_2$, hence, 
		no red is drawn. The black dot corresponds to the shrinking of $\mathcal{K}_\xi$ to a single point at the threshold value $\xi^*_{2}$.}
		\label{fig:Stackheat2}
\end{figure}

At the critical value $\xi^*_2$, the region $\mathcal{K}_\xi$ 
has shrunk to a single point. Clearly, this 
is a zero of the Routh-Hurwitz criterion, moreover, it is a 
local minimum both with respect to $\alpha$ and $\nu$. Hence, 
we look for $(\nu, \alpha, \xi)$ solving 
\begin{equation*}
\left[
\def\arraystretch{1.3}
\begin{array}{c}
    ~ ~ y_\nu(\alpha,\xi)  \\
    \frac{\partial}{\partial \alpha}y_\nu(\alpha,\xi) \\
    \frac{\partial}{\partial \nu} y_\nu(\alpha,\xi)\\
\end{array}
\right] = 
\left[
\def\arraystretch{1.3}
\begin{array}{c}
    0\\
    0\\
    0\\
\end{array}
\right],
\end{equation*}
leading to $\xi^*_2 \approx 9.19845 \times 10^{-3}$. 
For larger relative infectivity,  
{i.e.} $\xi>\xi_2^*$, there is no region 
of instability, thus $\mathcal{K}_\xi = \emptyset$, that is,  
the EE is LAS for all $(\alpha, \nu)$. 
The localized 
transition is visualized in 
Figure~\ref{fig:Stackheat2}.

Note that we did not investigate the 
dependence of these phenomena, 
and of the corresponding threshold values, 
on the other parameters 
fixed in \eqref{eq:parametrization}.

\subsection{Numerical bifurcation examples}
\label{sec:matcont}
In this section, we present numerical examples of one parameter $(\alpha)$ and 
two parameter $(\alpha, \nu)$ bifurcations of the endemic equilibria branch using \matcont { }\cite{MatCont_article}. An identical analysis we carried out in \cite{Richmond1} for an SIRWS system, therefore here we show some interesting examples to highlight the dynamics in the presence of the $J$ compartment. 

\begin{figure}[H]
	\centering
		\includegraphics[scale=0.8]{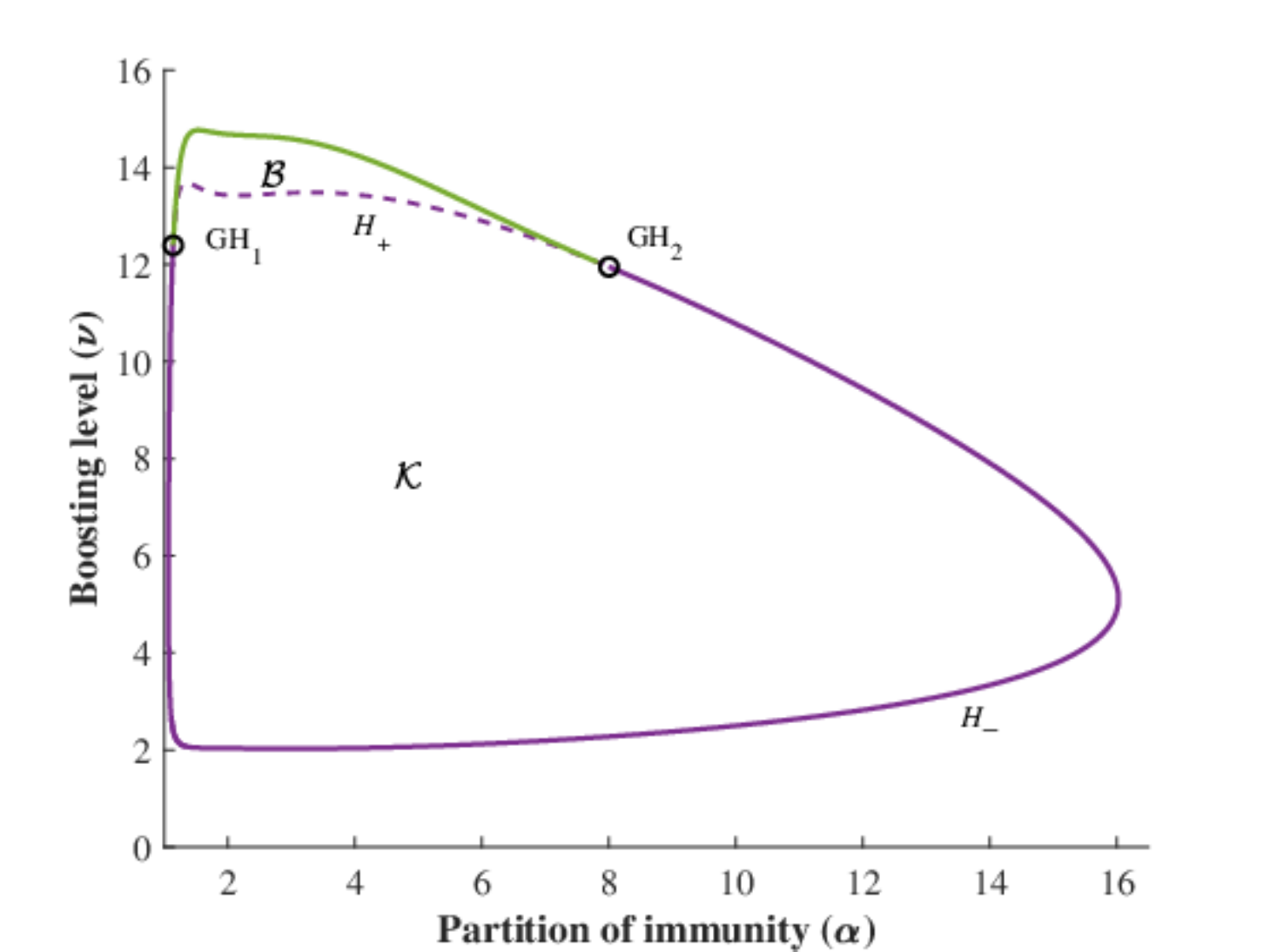}
		\caption{Two-parameter bifurcation diagram in the $(\alpha,\nu)$-plane, with $\xi=10^{-5}$.}
		\label{fig:NEWBIF}
\end{figure}

We briefly summarize the dynamics on the two parameter $(\alpha,\nu)$ bifurcation 
diagram when $\xi=10^{-5}<\xi_1^*$, see 
Figure~\ref{fig:NEWBIF}. The instability region $\mathcal K\,(\, =\mathcal{K}_\xi$ for fixed $\xi$) is enclosed by the purple-colored Hopf curve, which is continuous when supercritical (called $H_{-}$) and dashed when subcritical (called $H_{+}$). 
The two generalized Hopf points $\gh_1$ and 
$\gh_2 $, mark the parameter values where the Hopf bifurcation
changes from supercritical to subcritical. Note that these points now possess different $\nu$ coordinates as opposed to the limiting case in Figure~\ref{fig:heatmap-previous}. The branch of the limit points of periodic cycles appears in green, which together with the dashed purple curve $H_{+}$ enclose a bistability region $\mathcal{B}$, where there exists a stable periodic solution 
alongside the LAS endemic equilibrium. 

\begin{figure}[H]
\centering
\begin{subfigure}[b]{0.75\textwidth}
   \includegraphics[width=1\linewidth]{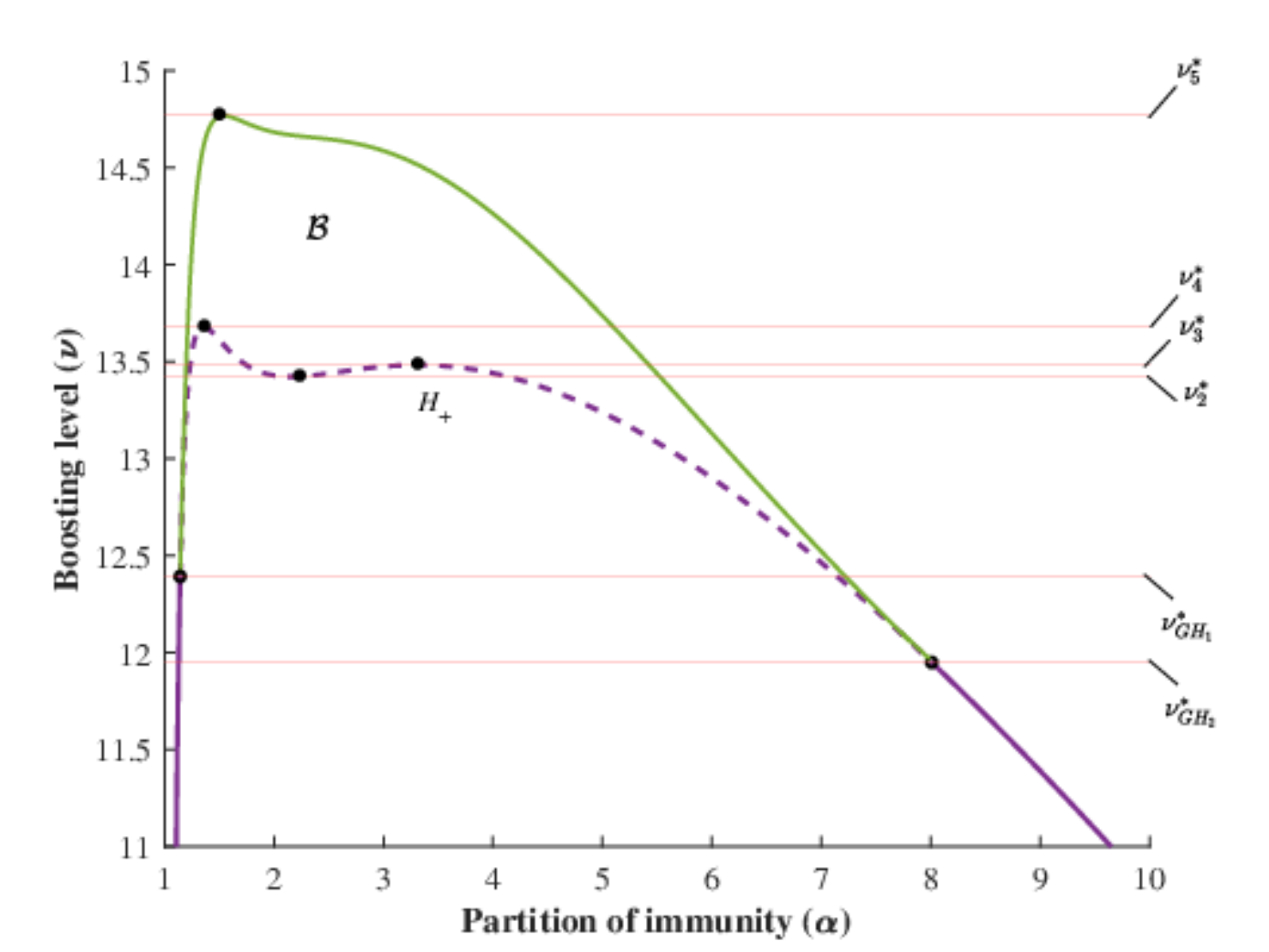}
   \caption{}
   \label{fig:Ng1} 
\end{subfigure}
\begin{subfigure}[b]{0.75\textwidth}
   \includegraphics[width=1\linewidth]{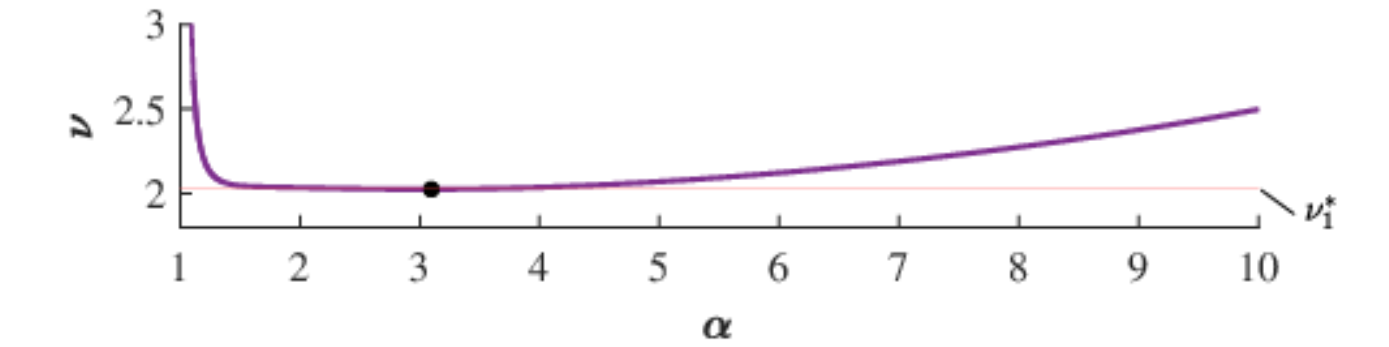}
   \caption{}
   \label{fig:Ng2}
\end{subfigure}
\caption{Two-parameter bifurcation diagram in the $(\alpha,\nu)$-plane, with $\xi=10^{-5}$ and critical $\nu$ values.}
\label{fig:NEWBIF_nu_stars}
\end{figure}

Let us now examine the bifurcation diagram in more detail  
over regions,
characterized by various levels of boosting rate $\nu$,
where the dynamics is similar, see Figure~\ref{fig:NEWBIF_nu_stars} for such partition and Table~\ref{tab:nu_stars} for the critical boosting values.

\begin{table}[H]
    \centering
    {\begin{tabular}{@{}lrrrrrr@{}} \toprule
         $\nu^*_{1}$ \qquad~
       & $\nu^*_{GH_2}$ &\qquad~ $\nu^*_{GH_1}$ &\qquad ~  $\nu^*_{2}$ &\qquad ~ $\nu^*_{3}$ &\qquad ~ $\nu^*_{4}$&$\nu^*_{5}$\\ 
         \midrule
         $2.0248$ 
        & $11.9494$& \qquad ~  $12.3922$ &$13.42$ & $13.48$ & $13.6785$& $14.7675$\\
         \bottomrule
    \end{tabular}}
    \caption{Approximate critical boosting values $(\nu^*_{k})$ using the parametrization \eqref{eq:parametrization} and fixing $\xi = 10^{-5}$ as in Figure~\ref{fig:NEWBIF_nu_stars}.}
    \label{tab:nu_stars}
\end{table}

In all bifurcation plots that follow, the endemic equilibria branch (particularly the $I$ and $J$ components) is marked with black curve, solid when LAS and dashed when unstable. Red and blue curves represent branches of stable and unstable limit cycles, respectively, and Hopf bifurcation points are marked with purple dots.

Below we describe the situation for various ranges of $\nu$.

\paragraph{Boosting: $\nu < \nu_1^*$} The system has a stable point attractor for all $\alpha > 1$.

\begin{figure}[H]
     \centering
     \subfloat[]{\includegraphics[scale=0.7]{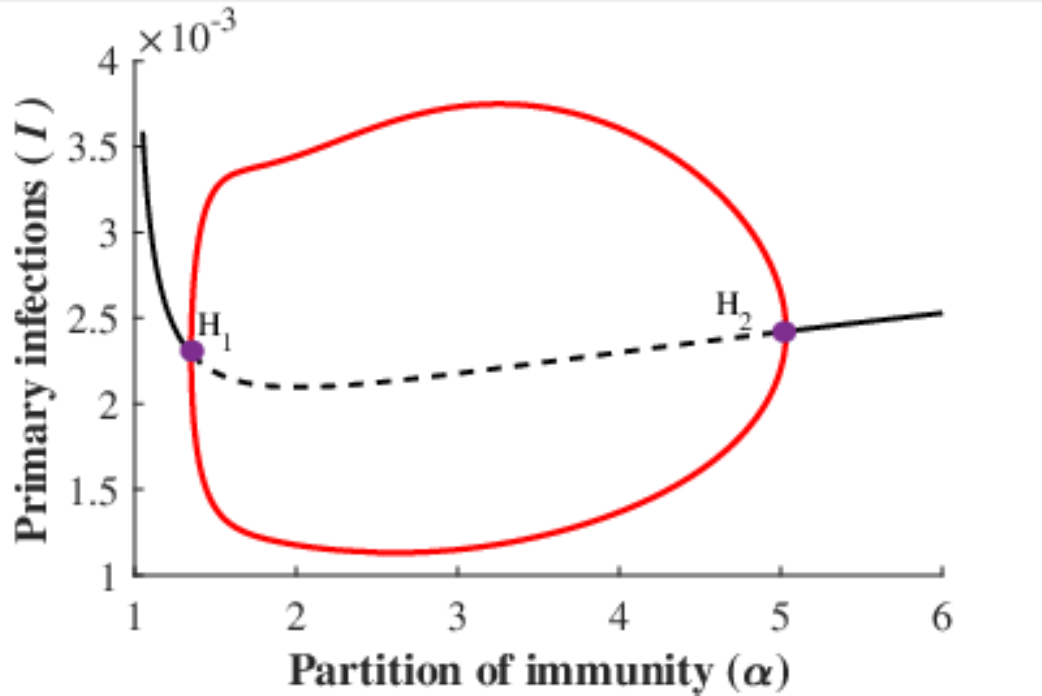}\label{fig:biff1}}
     \subfloat[]{\includegraphics[scale=0.7]{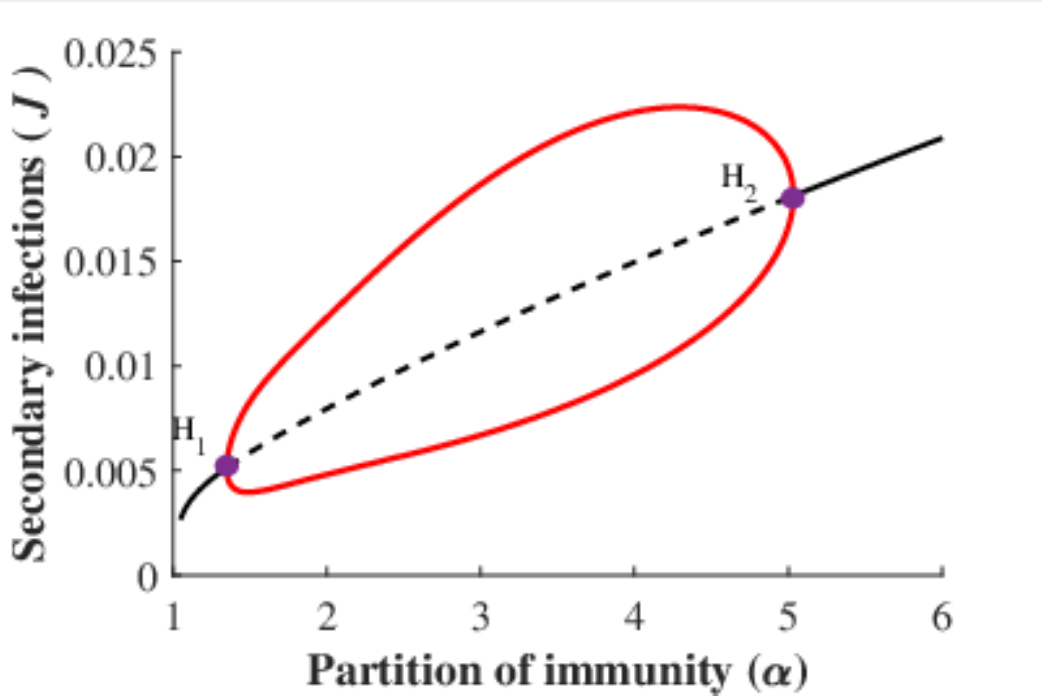}\label{fig:biff2}}
     \caption{One-parameter bifurcation diagram with $\xi=10^{-5}$ and $\nu=2.07$, (a) primary and (b) secondary infections.}
     \label{fig:biffff1}
\end{figure}

\paragraph{Boosting: $\nu_1^* < \nu < \nu_{GH_2}^*$}
There are two supercritical Hopf bifurcation points on the endemic equilibria branch, 
see the lower inset in Figure~\ref{fig:RH-Large-Left1}. Continuation of limit cycles with respect to $\alpha$ starting from two Hopf bifurcation points, $\h_1$ and $\h_2$, forms an endemic bubble, where the two branches of stable limit cycles coincide, see Figure~\ref{fig:biffff1}.

\paragraph{Boosting: $\nu_{GH_2}^* < \nu < \nu_{GH_1}^*$}
As $\nu$ continues to grow in the two-parameter plane in Figure~\ref{fig:NEWBIF_nu_stars} the generalized Hopf point $\gh_2$ appears, which separates branches of sub- and supercritical Hopf bifurcations. The stable limit cycles survive when we enter the region $\mathcal{B}$. Crossing the subcritical Hopf boundary $H_+$ leads to an additional unstable cycle inside the first one, while the equilibrium regains its stability. Two cycles of opposite stability exist inside the bistable region $\mathcal{B}$ and disappear at the green curve. 

\begin{figure}[H]
	\centering
		\subfloat[]{\includegraphics[scale=0.72]{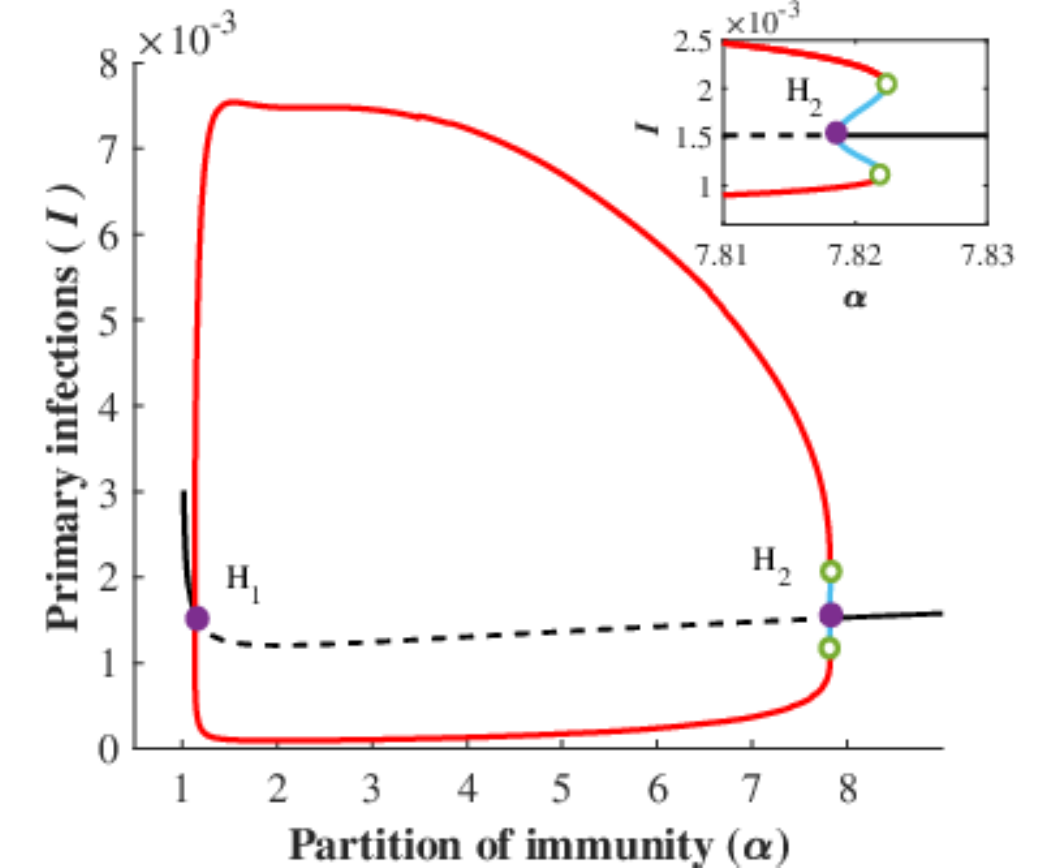}}
		\subfloat[]{\includegraphics[scale=0.72]{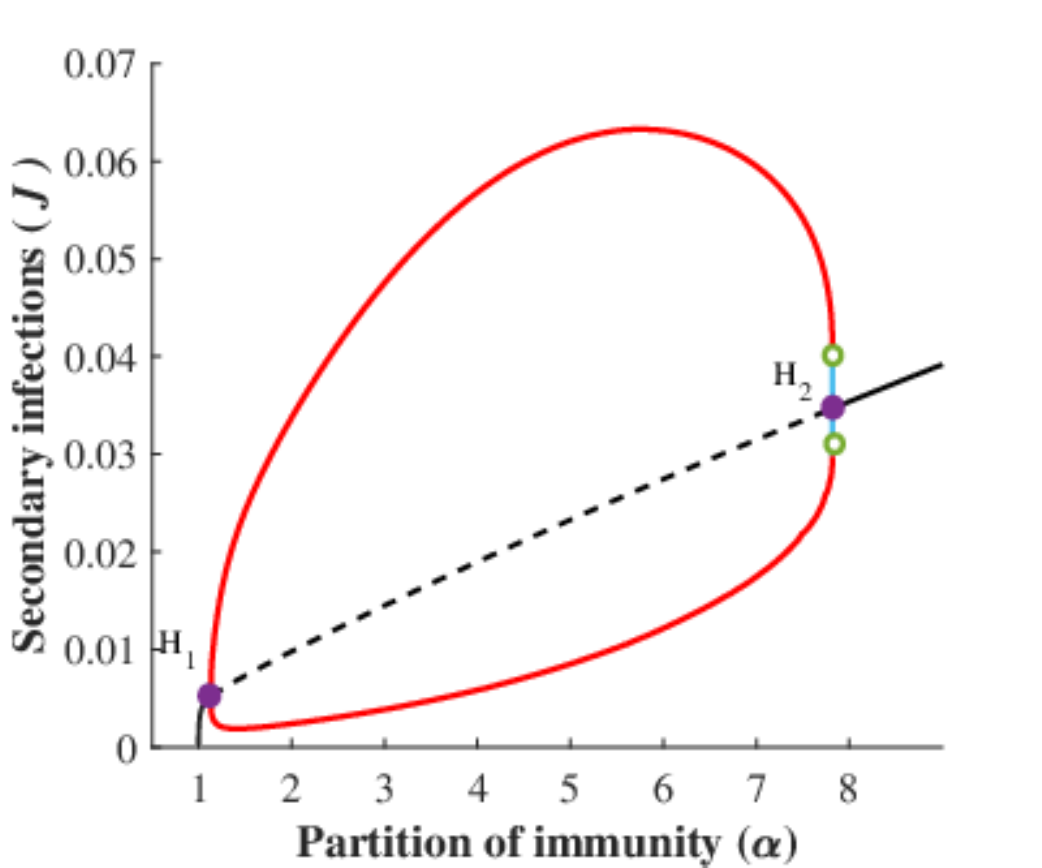}}
		\caption{One-parameter bifurcation diagram with $\xi=10^{-5}$ and $\nu=12.05 $, (a) primary and (b) secondary infections.}
		\label{fig:nu_12_05}
\end{figure}

Let us fix $\nu$ in this boosting region. Then Figure~\ref{fig:nu_12_05} shows a typical bifurcation with respect to $\alpha$. Observe here the small $\alpha$-parameter range of bistability where the EE and the larger amplitude periodic solution are both stable. The points marked with green circle are limit points of periodic orbits. The stable and unstable cycles collide and disappear on the green curve in Figure~\ref{fig:NEWBIF_nu_stars}, corresponding to a fold bifurcation of cycles. 

\begin{figure}[H]
	\centering
		\subfloat[]{\includegraphics[scale=0.72]{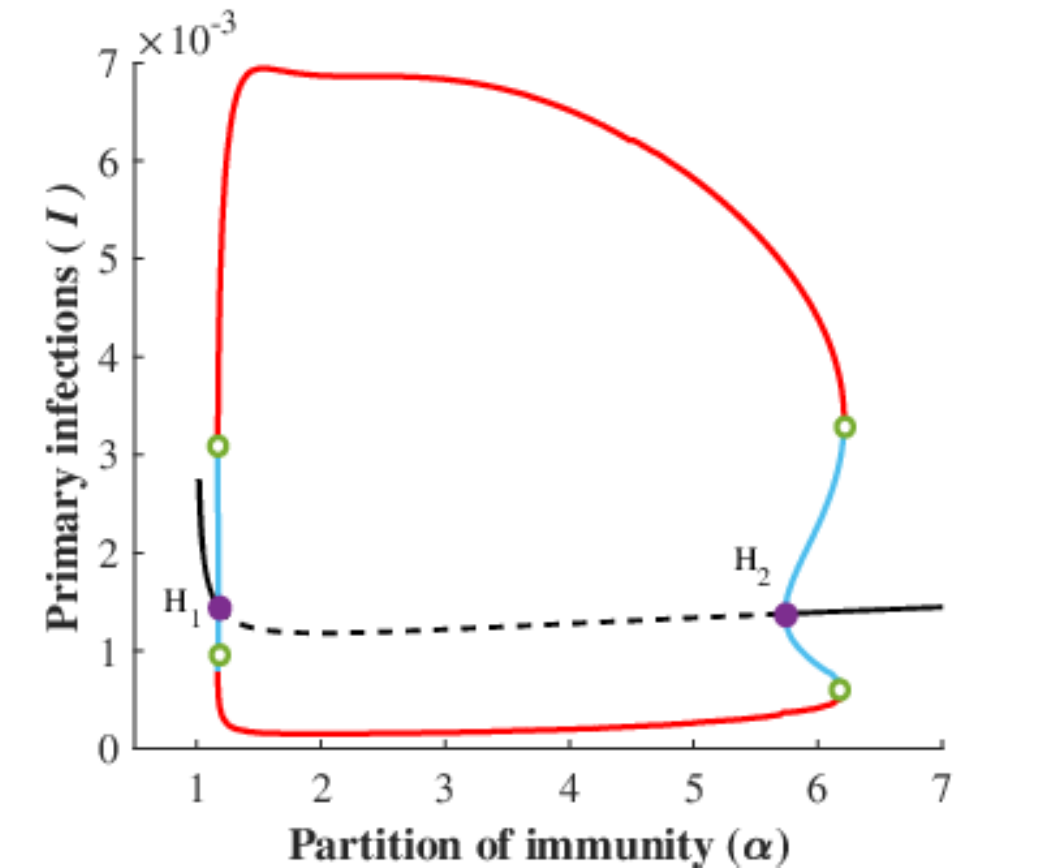}}
		\subfloat[]{\includegraphics[scale=0.72]{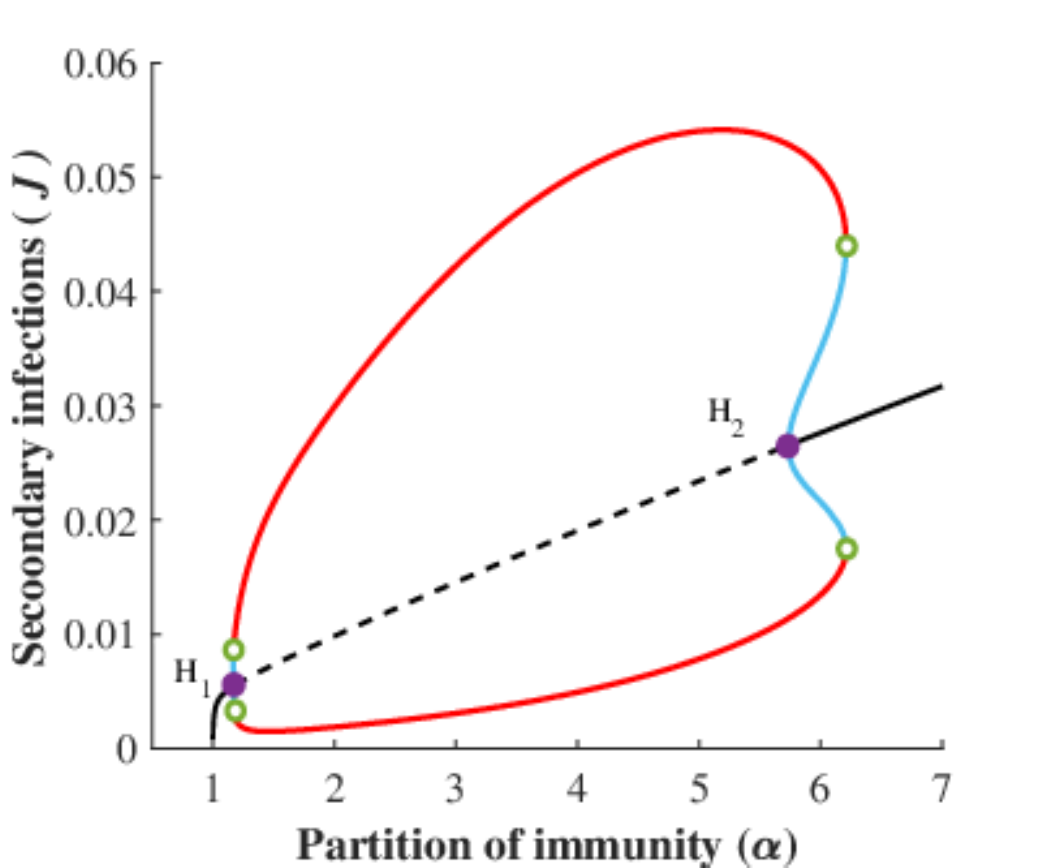}}
		\caption{One-parameter bifurcation diagram with $\xi=10^{-5}$ and $\nu=13$, (a) primary and (b) secondary infections.}
		\label{fig:nu_13}
\end{figure}

\paragraph{Boosting: $\nu_{GH_1}^* < \nu < \nu_2^*$}
In this boosting range, as we passed $\gh_1$, the Hopf curve changed to subcritical. Figure~\ref{fig:nu_13} confirms the appearance of two subcritical Hopf bifurcations on the equilibria branch, then again a fold bifurcation of cycles occurs (marked with green circles), resulting in two small $\alpha$-parameter intervals of bistability. 

\begin{figure}[H]
     \centering
     \subfloat[]{\includegraphics[scale=0.72]{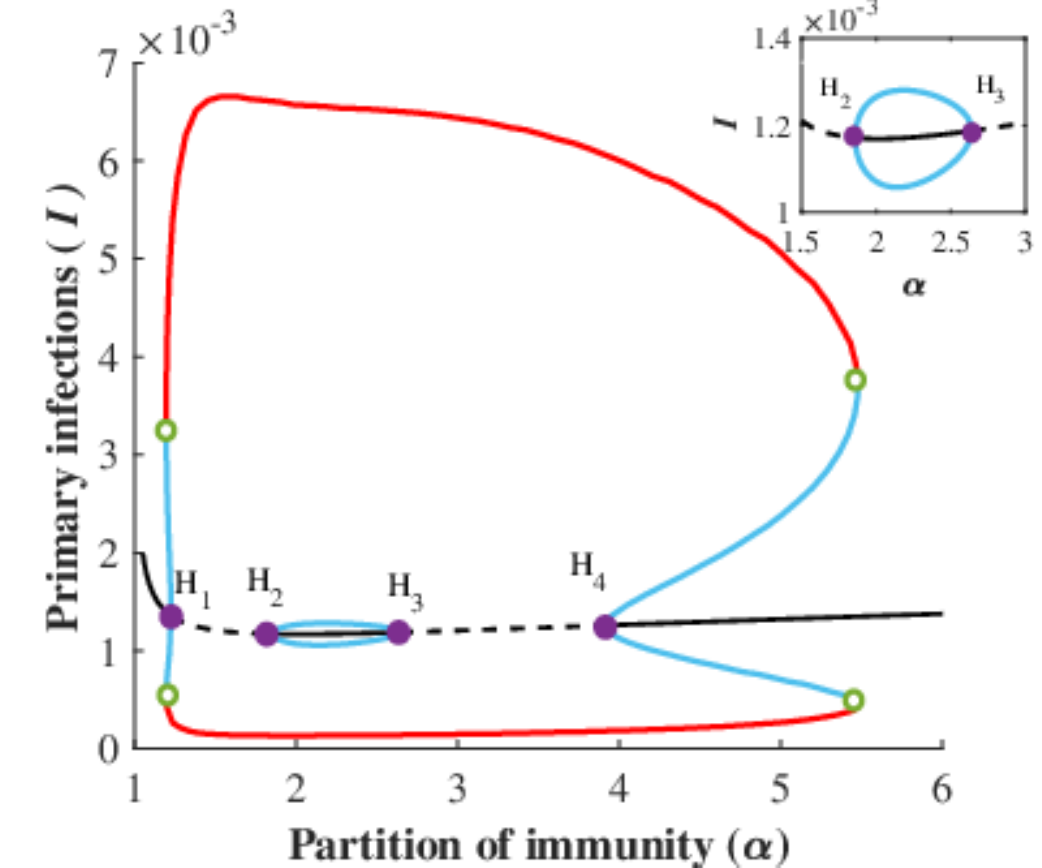}\label{fig:Amoah}}
     \subfloat[]{\includegraphics[scale=0.72]{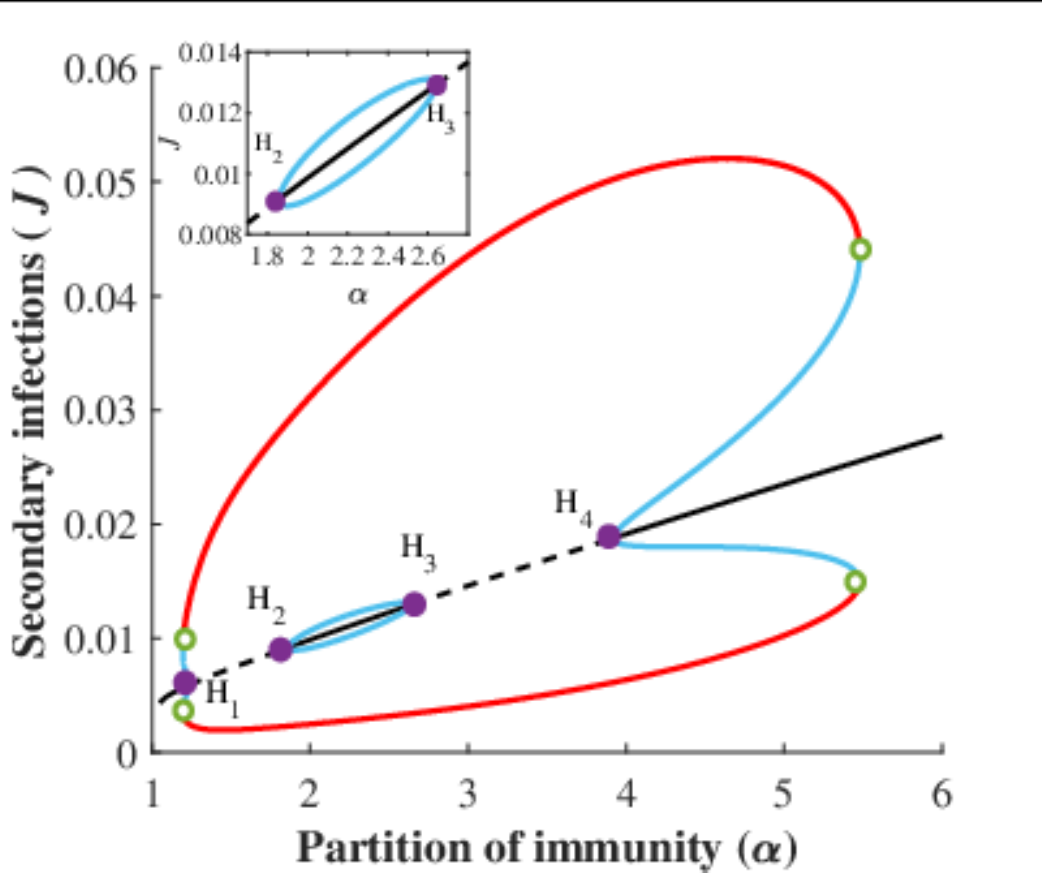}\label{fig:bubble1}}
     \caption{One-parameter bifurcation diagram with $\xi=10^{-5}$ and $\nu=13.45$, (a) primary and (b) secondary infections.}
     \label{fig:Amoah1}
\end{figure}

\paragraph{Boosting: $\nu_2^* < \nu < \nu_3^*$}
In this region, we can observe how the shape of the 
Hopf curve $H_+$ that bounds 
the set $\mathcal{K}_\xi$ influences the number of stability switches of
the EE. In Figure~\ref{fig:Amoah1}, the bifurcation diagram shows the existence of four subcritical Hopf bifurcation points. Here, a small bubble appears inside the region of stable oscillations, which leads to an additional bistable region compared to the previous case.  
When we increase the boosting parameter but still being in this region, then the Hopf points $\h_1$ and $\h_2$ as well as $\h_3$ and $\h_4$ move closer to each other, resulting in larger bistability regions, see also Figure~\ref{fig:NEWBIF_nu_stars}.

\begin{figure}[H]
	\centering
		\subfloat[]{\includegraphics[scale=0.72]{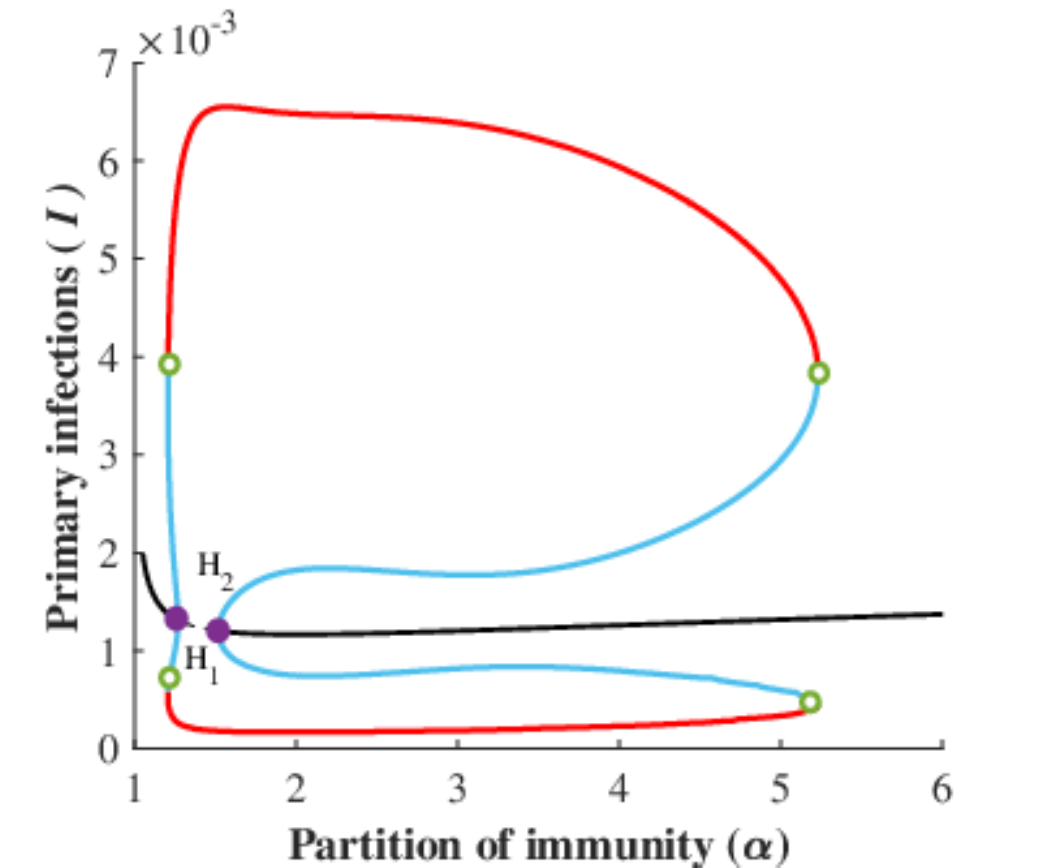}}
		\subfloat[]{\includegraphics[scale=0.72]{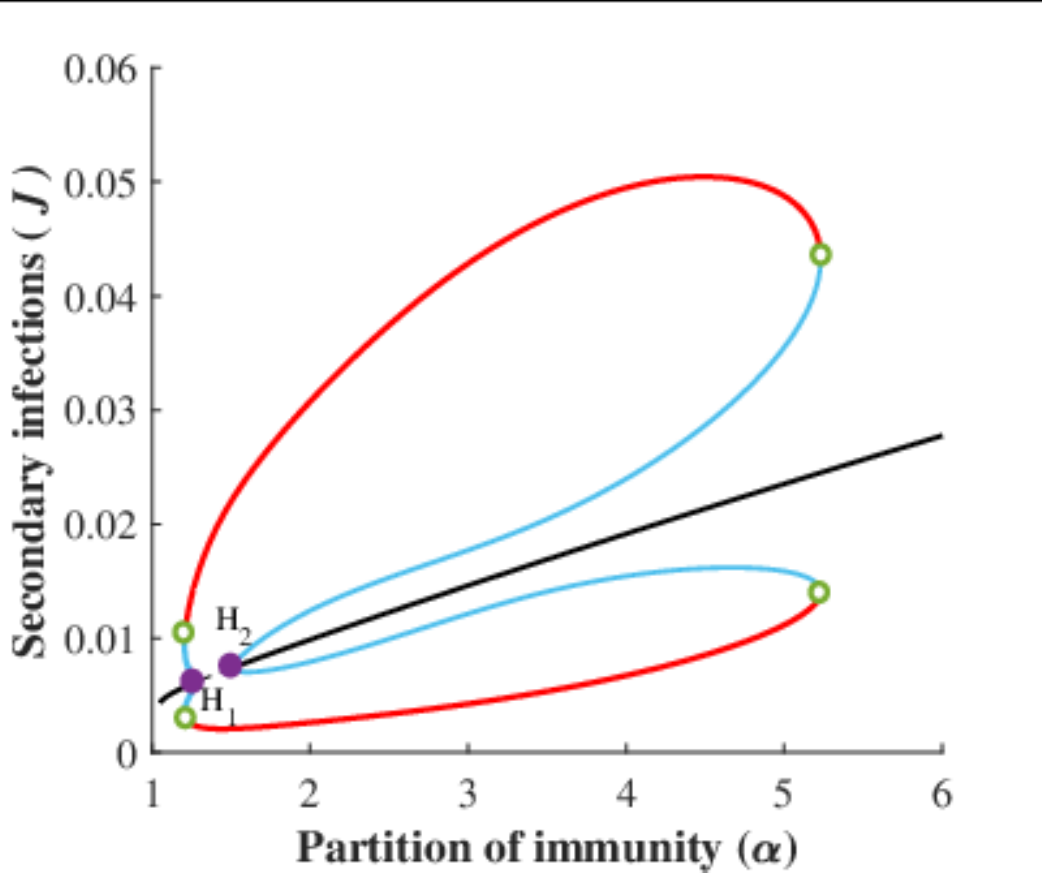}}
		\caption{One-parameter bifurcation diagram with $\xi=10^{-5}$ and $\nu=13.6$, (a) primary and (b) secondary infections.}
		\label{fig:prim1}
\end{figure}

\paragraph{Boosting: $\nu_3^* < \nu < \nu_4^*$}
Here, the two Hopf points $\h_3$ and $\h_4$ seen in the region before collided and disappeared, see Figure~\ref{fig:prim1}. The dynamics is similar to Figure~\ref{fig:nu_13} but the boosting values in this range lead to much larger bistability regions.

\begin{figure}[H]
	\centering
		\subfloat[]{\includegraphics[scale=0.72]{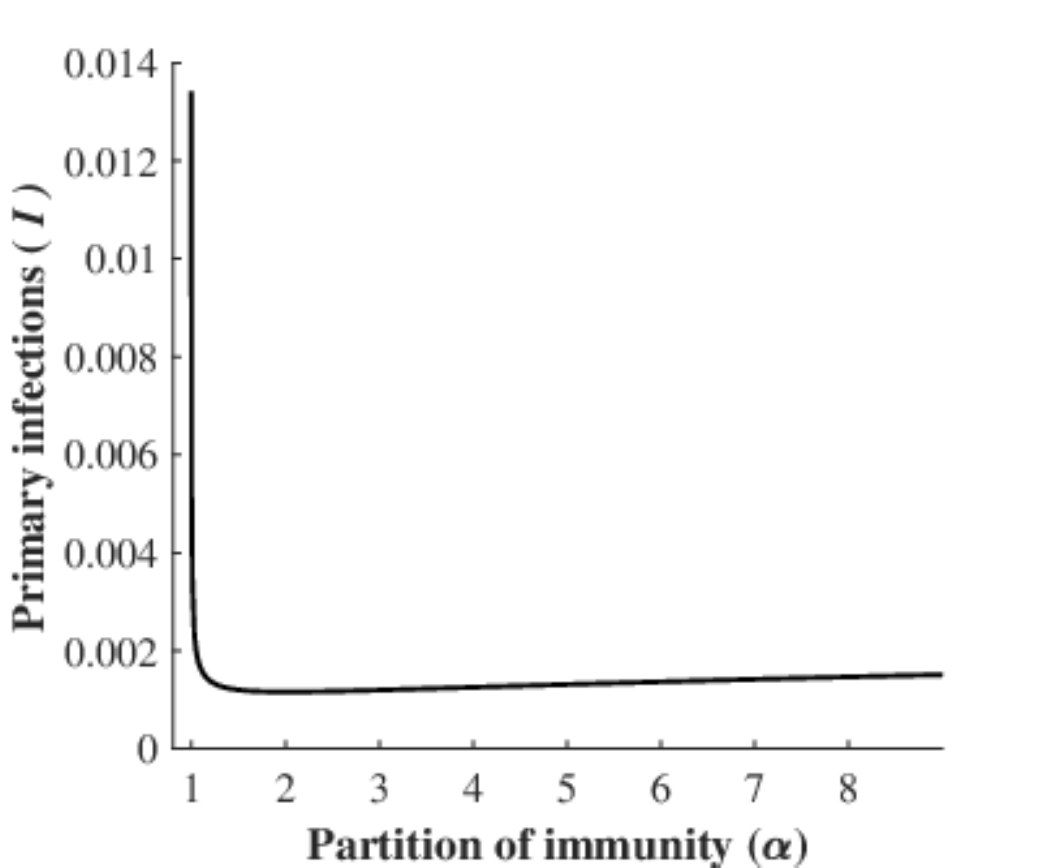}}
		\subfloat[]{\includegraphics[scale=0.72]{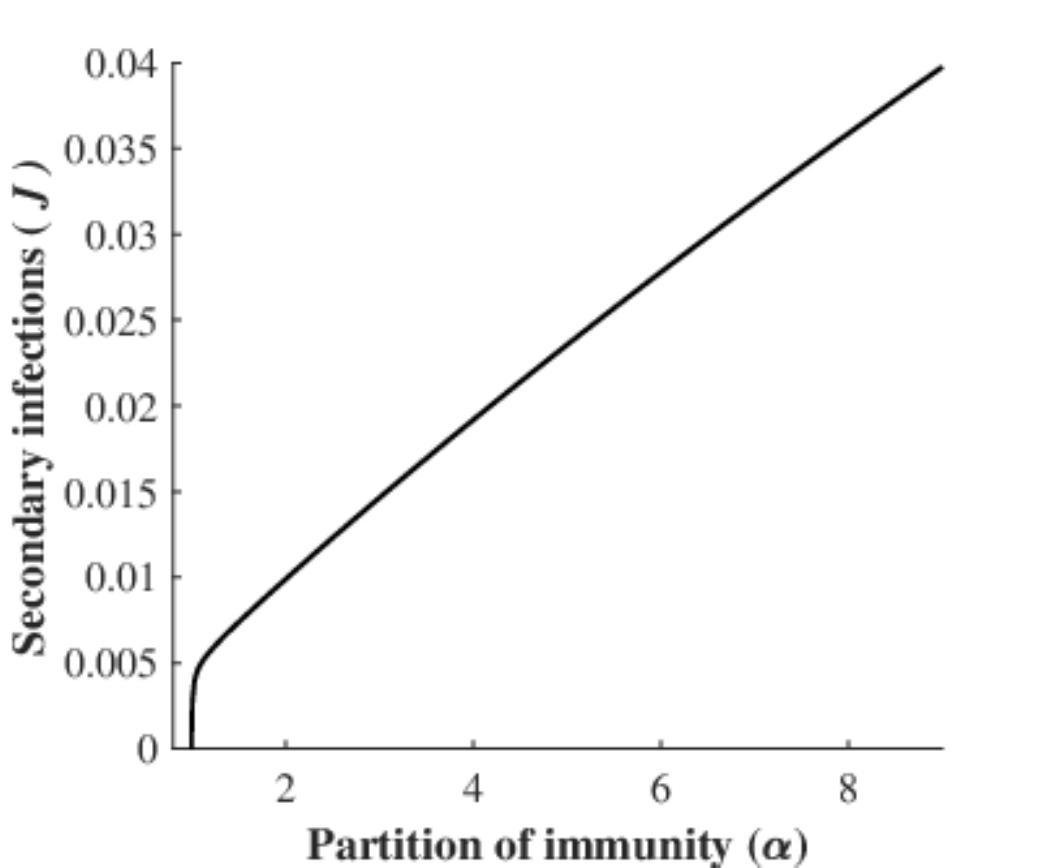}}
		\caption{One-parameter bifurcation diagram with $\xi=10^{-5}$ and $\nu=13.8$, (a) primary and (b) secondary infections.}
		\label{fig:nu_13_8}
\end{figure}

\paragraph{Boosting: $\nu_4^* < \nu < \nu_5^*$}
Although, we are in the bistability region in the two-parameter bifurcation plot, we do not cross any Hopf curve, hence the numerical continuation method finds a stable equilibrium branch, see Fig~\ref{fig:nu_13_8}.

\paragraph{Boosting: $\nu_5^* < \nu$}
The system has a stable point attractor for all $\alpha > 1$.

\begin{figure}[htb]
    \centering 
\begin{subfigure}{0.30\textwidth}
  \includegraphics[width=\linewidth]{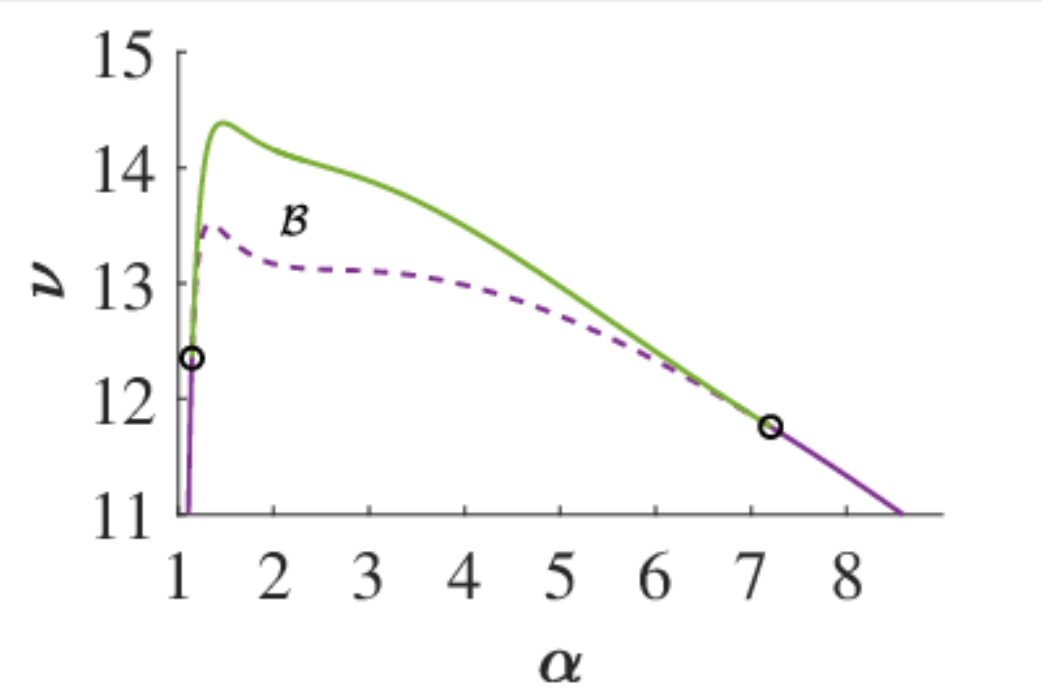}
  \caption{$\xi=5\times 10^{-5}$}
  \label{fig:bshrink1}
\end{subfigure}\hfil 
\begin{subfigure}{0.30\textwidth}
  \includegraphics[width=\linewidth]{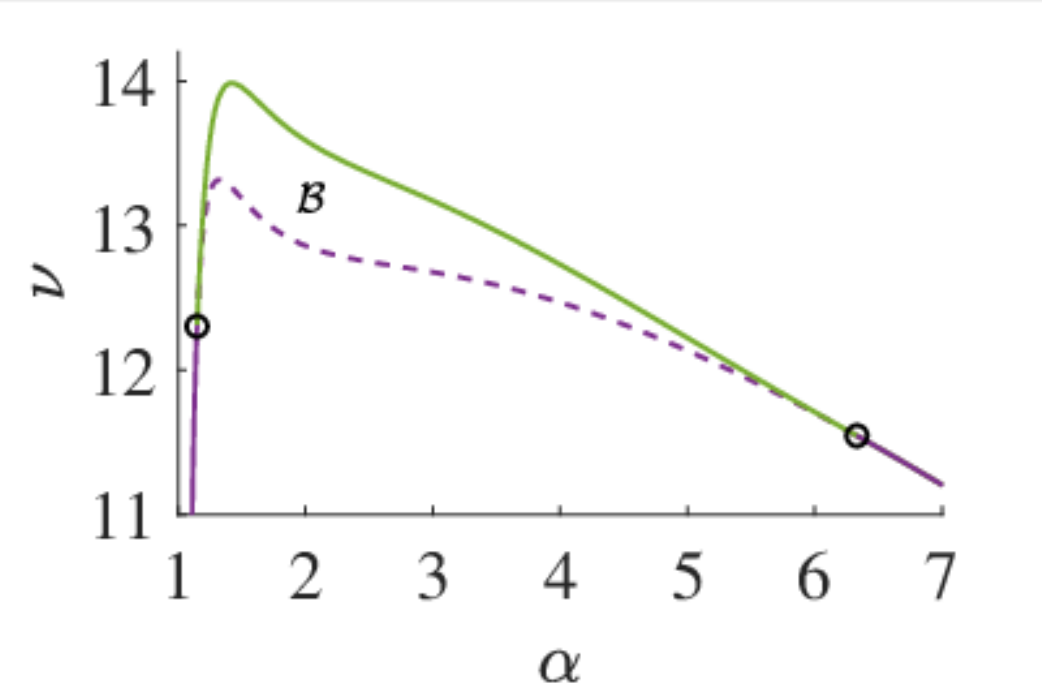}
  \caption{$\xi=1\times 10^{-4}$}
  \label{fig:bshrink2}
\end{subfigure}\hfil 
\begin{subfigure}{0.30\textwidth}
  \includegraphics[width=\linewidth]{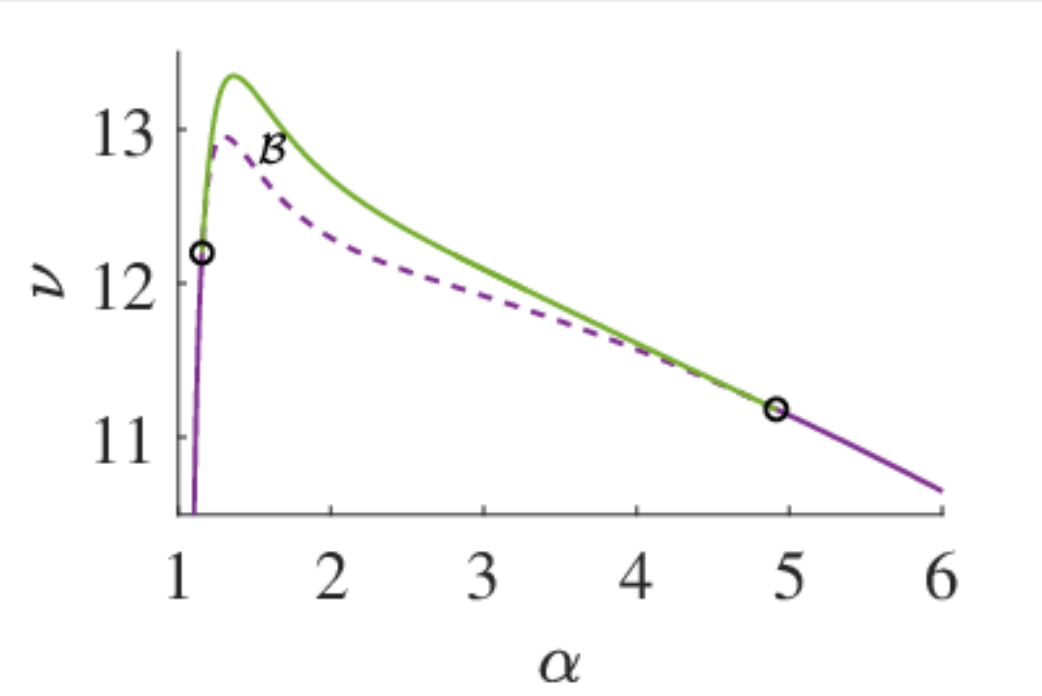}
  \caption{$\xi=2\times 10^{-4}$}
  \label{fig:bshrink3}
\end{subfigure}
\medskip
\medskip
\begin{subfigure}{0.30\textwidth}
  \includegraphics[width=\linewidth]{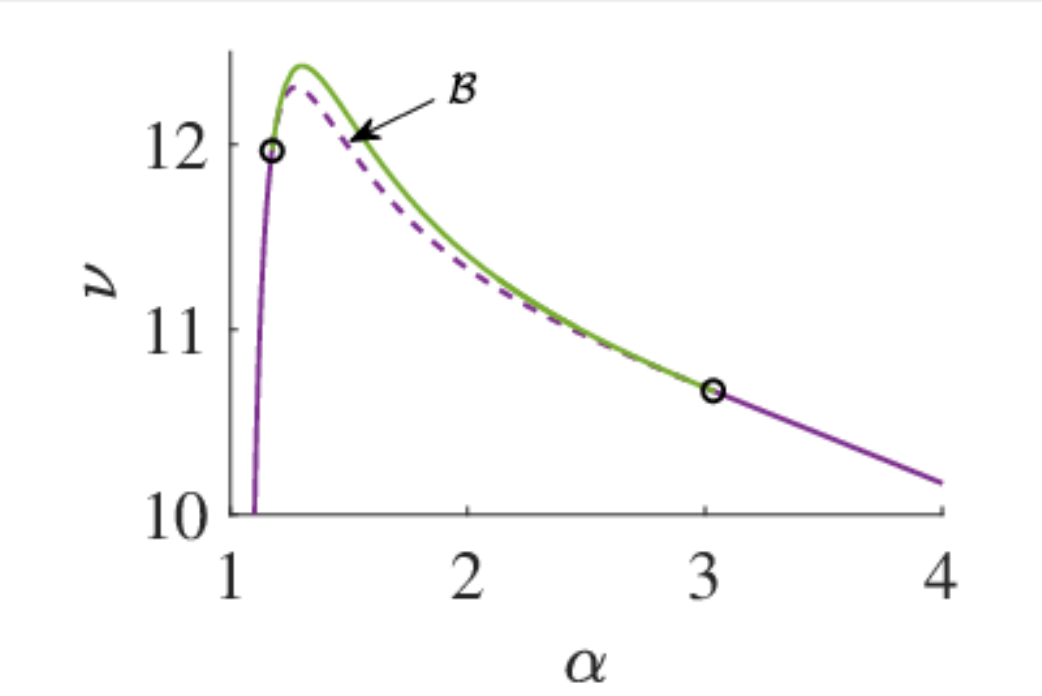}
  \caption{$\xi=4\times 10^{-4}$}
  \label{fig:bshrink4}
\end{subfigure}\hfil 
\begin{subfigure}{0.30\textwidth}
  \includegraphics[width=\linewidth]{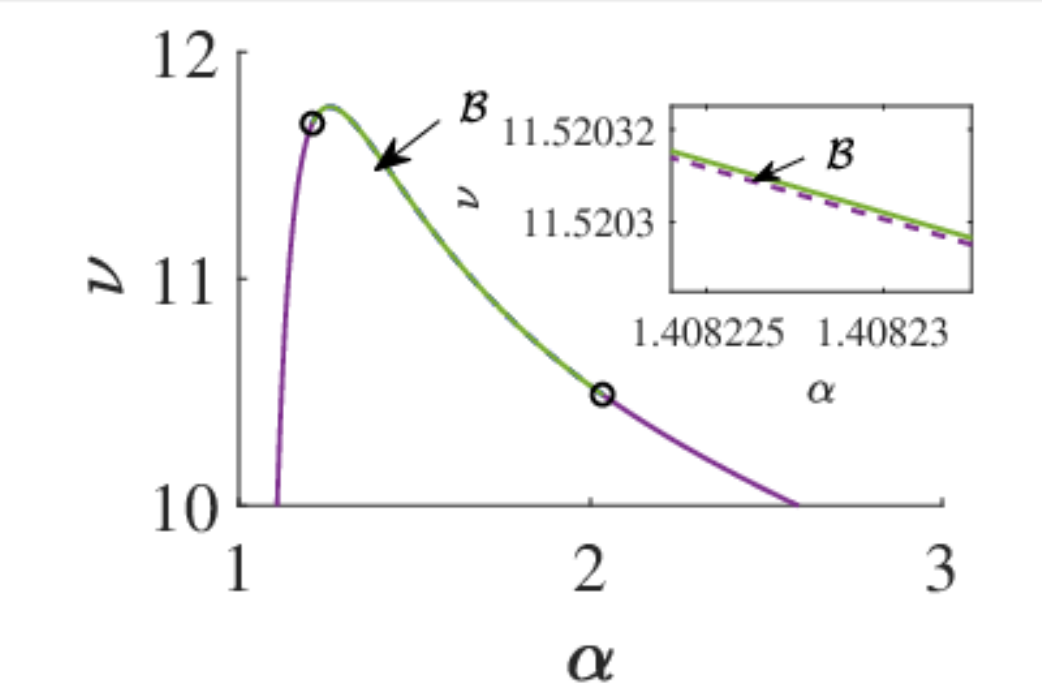}
  \caption{$\xi=6\times 10^{-4}$}
  \label{fig:bshrink5}
\end{subfigure}\hfil 
\begin{subfigure}{0.30\textwidth}
  \includegraphics[width=\linewidth]{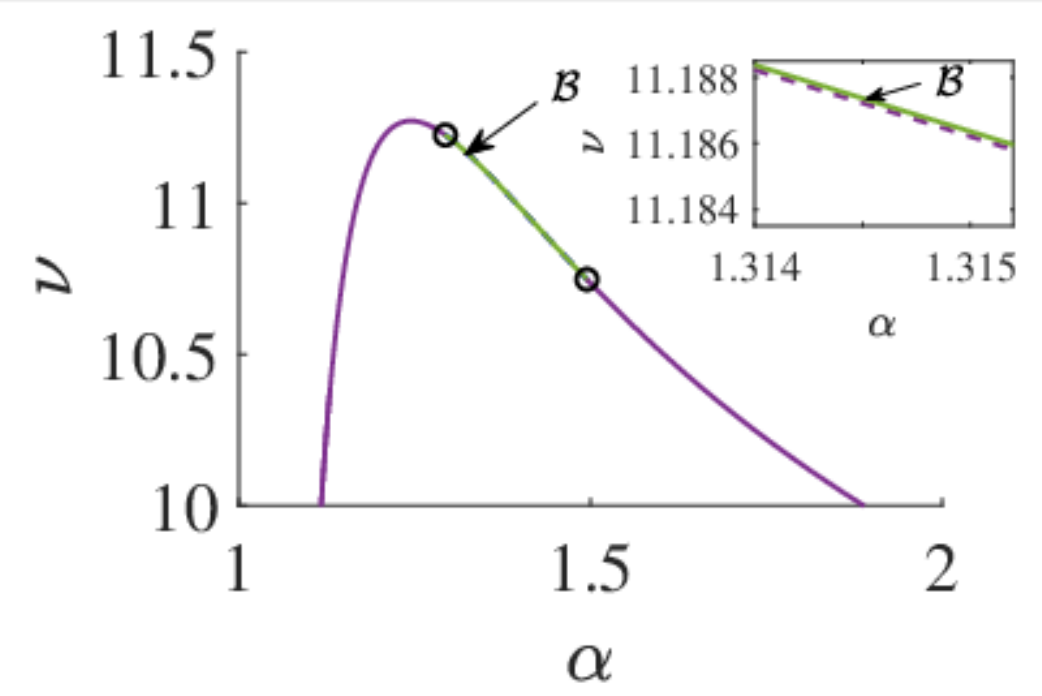}
  \caption{$\xi=8\times 10^{-4}$}
  \label{fig:bshrink6}
\end{subfigure}
\smallskip
\begin{subfigure}{0.30\textwidth}
  \includegraphics[width=\linewidth]{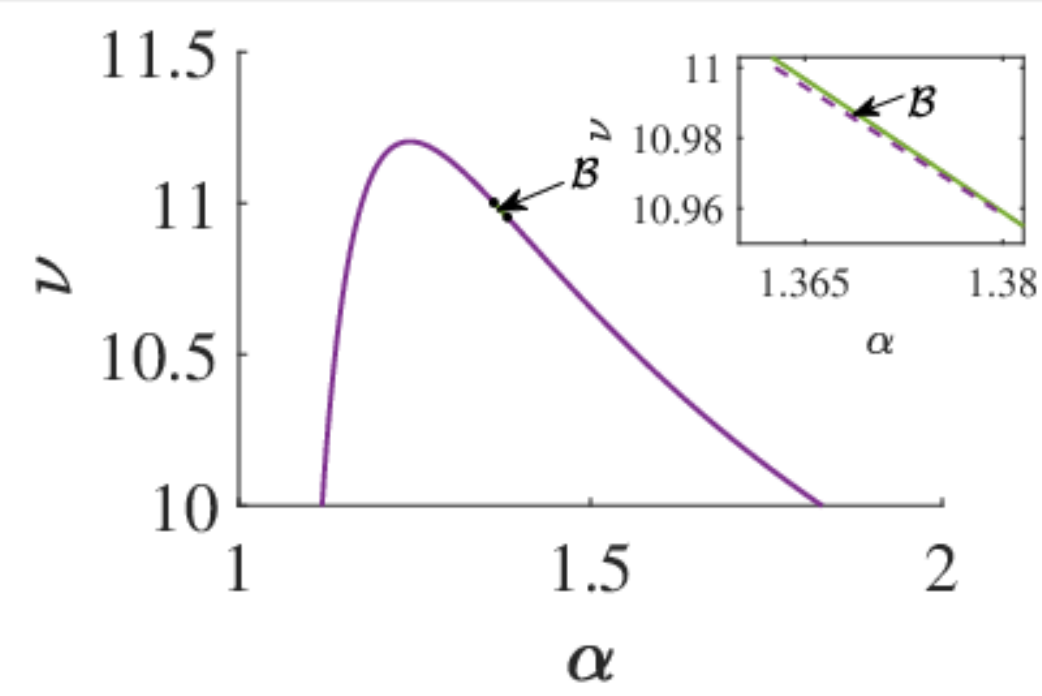}
  \caption{$\xi=8.3\times 10^{-3}$}
  \label{fig:bshrink7}
\end{subfigure}\hfil 
\begin{subfigure}{0.30\textwidth}
  \includegraphics[width=\linewidth]{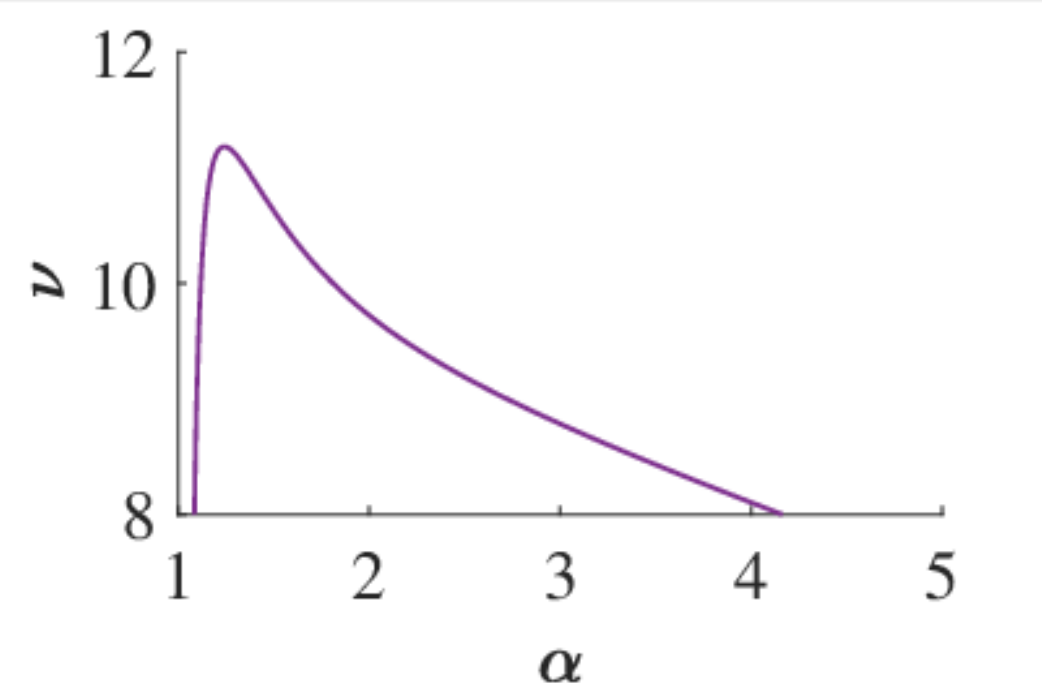}
  \caption{$\xi=8.4\times 10^{-3}$}
  \label{fig:bshrink8}
\end{subfigure}\hfil 
\caption{Bistabilty region ($\mathcal{B}$) on the two-parameter bifurcation diagram. The region is shrinking as demonstrated by figures a) - g) and has completely disappeared in figure h).}
\label{fig:B-Shrinks}
\end{figure}

\paragraph{Shrinking of the bistability region}
In Section~\ref{sec:EE-stability} we analyzed the shrinking of the instability region $\mathcal{K}_\xi$ as $\xi$ increases. As a consequence, the bistability region $\mathcal B$ becomes smaller, the generalized Hopf points move towards each other, then collide and disappear as illustrated in Figure~\ref{fig:B-Shrinks}. We did not localize further the threshold value $\xi^*_{\footnotesize \mathcal{B}} \in (8.3, 8.4) \times 10^{-3}$ at which this region disappears.

\section{Conclusion}
\label{sec:conclusion}
In this paper, we carried out combined analytical and numerical investigations of the $SIRWJS$ system with the presence of secondary infections and potentially asymmetric partitioning of the immune boosting period. As the model population is assumed to be constant, the system is inherently four dimensional resulting in rather 
complicated formulae describing the equilibria and their stability. The analysis presented in this manuscript is 
giving us novel insights into this complexity and a 
better understanding of the dynamics. We concluded an exact condition in the form of $\SuperCond$ determining the direction of 
the bifurcation at $\rzero = 1$, and showed that backward bifurcation is possible.
For $\rzero > 1$, we derived a numerically tractable Routh-Hurwitz stability criterion and carried out its sign analysis together with numerical continuation techniques. We observed rich and interesting dynamics in the $(\nu, \alpha, \xi)$-space that is varying the immune boosting rate, the partitioning of the boosting period, and the relative infectivity of secondary infections, where other disease parameters were set according to pertussis parameter values taken from the literature.   
Nevertheless, we note that most of the mathematically appealing phenomena occur for rather large boosting rate ($\nu$) and very small relative infectivity ($\xi$). 

Overall, our results highlight the potential complexities generated by the combination of waning and boosting of immunity, which makes the prediction of the long term dynamics rather challenging for diseases when these processes are relevant, such as COVID-19 or pertussis. 


\section*{Acknowledgement}
This research was completed in the National Laboratory for Health Security RRF-2.3.1-21-2022-00006. Additionally, the work was supported by the Ministry of Innovation and Technology of Hungary from the National Research, Development and Innovation Fund project no. TKP2021-NVA-09. In addition, F.B. and G.R. were supported by NKFIH grant numbers FK 138924, KKP 129877. M.P. was also supported by the Hungarian Scientific Research Fund grant numbers K129322 and SNN125119; F.B. was also supported by UNKP-22-5 and the Bolyai scholarship of the Hungarian Academy of Sciences.



\begin{thebibliography}{99}
\addcontentsline{toc}{section}{References}

\bibitem{general}
M.V. Barbarossa, G. Röst, 
Immuno-epidemiology of a population structured by immune status: 
a mathematical study of waning immunity and immune system boosting, 
J. Math. Biol.
71 (6) (2015) 1737--1770.\\
\url{https://doi.org/10.1007/s00285-015-0880-5}.\\
\url{https://pubmed.ncbi.nlm.nih.gov/25833186}.\\
\url{https://mathscinet.ams.org/mathscinet-getitem?mr=3419906}.

\bibitem{carlsson}
R.M. Carlsson, L.M. Childs, Z. Feng, J.W. Glasser, J.M. Heffernan, J. Li, G. Röst, 
Modeling the waning and boosting of immunity from infection or vaccination, 
J. Theor. Biol. 497 (2020) 110265. \\ 
\url{https://doi.org/10.1016/j.jtbi.2020.110265}. \\
\url{https://pubmed.ncbi.nlm.nih.gov/32272134}.

\bibitem{castillo2004dynamical}
C. Castillo-Chavez, B. Song, 
Dynamical models of tuberculosis and their applications, 
Math. Biosci. Eng. 
1 (2) (2004) 361--404.\\
\url{https://doi.org/10.3934/mbe.2004.1.361}.\\
\url{https://pubmed.ncbi.nlm.nih.gov/20369977}.\\
\url{https://mathscinet.ams.org/mathscinet-getitem?mr=2130673}.

\bibitem{childs}
L. Childs, D.W. Dick, Z. Feng, J.M. Heffernan, J. Li, G. Röst, Modeling waning and boosting of {COVID-19} in Canada with vaccination,  Epidemics, 39 (2002) 100583.
\url{https://doi.org/10.1016/j.epidem.2022.100583}. \\ 
\url{https://pubmed.ncbi.nlm.nih.gov/35665614}.

\bibitem{dafilis2012influence}
M.P. Dafilis, F. Frascoli, J.G. Wood, J.M. McCaw, The influence of increasing life expectancy on the dynamics of {SIRS}
systems with immune boosting, 
The ANZIAM Journal 
54 (1-2) (2012) 50--63.\\ 
\url{https://doi.org/10.1017/S1446181113000023}.\\
\url{https://mathscinet.ams.org/mathscinet-getitem?mr=3066288}.

\bibitem{MatCont_article}
A. Dhooge, W. Govaerts, Y.A. Kuznetsov, MATCONT: a Matlab package for numerical bifurcation analysis of ODEs, 
SIGSAM Bull. 38 (1) (2004) 21--22.\\
\url{https://doi.org/10.1145/980175.980184}.\\
\url{https://mathscinet.ams.org/mathscinet-getitem?mr=2000880}.

\bibitem{diekmann}
O. Diekmann,  J.A.P. Heesterbeek, M.G. Roberts, 
The construction of next-generation matrices for
compartmental epidemic models, J. R. Soc. Interface 7 (47) (2010) 873--885.\\
\url{https://doi.org/10.1098/rsif.2009.0386}. \\
\url{https://pubmed.ncbi.nlm.nih.gov/19892718}.

\bibitem{Heidecke2021}
J. Heidecke, M.V. Barbarossa,
{When Ideas Go Viral---Complex Bifurcations in a Two-Stage Transmission Model}, 
In: Mondaini, R.P. (eds) Trends in Biomathematics: Chaos and Control in Epidemics, Ecosystems, and Cells. BIOMAT 2020. Springer, Cham. \\
\url{https://doi.org/10.1007/978-3-030-73241-7_14}.\\
\url{https://mathscinet.ams.org/mathscinet-getitem?mr=4306495}.

\bibitem{KATRIEL2019}
G. Katriel, 
{The dynamics of two-stage contagion},
{ Chaos, Solitons \& Fractals: X}, 2 (2019) 100010.\\ 
\url{https://doi.org/10.1016/j.csfx.2019.100010}.

\bibitem{lavine2011natural}
J.S. Lavine, A.A. King, O.N. Bj{\o}rnstad,  
{ Natural immune boosting in pertussis dynamics 
and the potential for long-term vaccine failure}, 
PNAS 
108 (17) (2011) 7259--7264.\\
\url{https://doi.org/10.1073/pnas.1014394108}.\\
\url{https://pubmed.ncbi.nlm.nih.gov/21422281}.

\bibitem{leung2018infection}
T. Leung, P.T. Campbell, B.D. Hughes, F. Frascoli, 
J.M. McCaw, 
{ Infection-acquired versus vaccine-acquired immunity in an {SIRWS} model}, 
Infect. Dis. Model. 
3 (2018) 118--135.\\ 
\url{https://doi.org/10.1016/j.idm.2018.06.002}.\\
\url{https://pubmed.ncbi.nlm.nih.gov/30839933}.

\bibitem{liu}
M. Liu, E. Liz, G. R\"ost, 
{Endemic bubbles generated by delayed behavioral response: global stability and bifurcation switches in an SIS model}, 
SIAM J. Appl. Math. 75 (1) (2015) 75--91.\\
\url{https://doi.org/10.1137/140972652}.\\
\url{https://mathscinet.ams.org/mathscinet-getitem?mr=3299143}.

\bibitem{murray2007mathematical}J.D. Murray, 
{Mathematical Biology: I. An Introduction. Interdisciplinary Applied Mathematics, 17}, 
Springer-Verlag, 
New York, New York, 
2002.\\ 
\url{https://doi.org/10.1007/b98868}. \\ 
\url{https://mathscinet.ams.org/mathscinet-getitem?mr=1908418}.


\bibitem{Richmond1}
R. Opoku-Sarkodie, F.A. Bartha, M. Polner, G. Röst,
{Dynamics of an SIRWS model with waning of immunity and varying immune boosting,} 	J. Biol. Dyn. 16 (1) (2022) 596--618.\\
\url{https://doi.org/10.1080/17513758.2022.2109766}.\\
\url{https://pubmed.ncbi.nlm.nih.gov/35943129}.\\
\url{https://mathscinet.ams.org/mathscinet-getitem?mr=4466048}.

\bibitem{routh1877treatise}
E.J. Routh, 
{ A treatise on the Stability of a given state of motion: 
particularly steady motion}, 
Macmillan and Co., 
London, 
1877. 

\bibitem{strube}
L.F. Strube, M. Walton, L.M. Childs, 
{ Role of repeat infection in the dynamics
of a simple model of waning and boosting immunity}, 
J. Biol. Syst. 
29 (2) (2021) 1--22.\\
\url{https://doi.org/10.1142/S0218339021400076}.\\
\url{https://mathscinet.ams.org/mathscinet-getitem?mr=4274350}.

\bibitem{github}Computer Algebra Codes, Github, 2023.\\ 
\url{https://github.com/epidelay/waning-boosting-epidemiological-models}.
\end{thebibliography}
\end{document}